\documentclass[sigconf, nonacm]{acmart}

\graphicspath{{Figures}}
\usepackage{hyperref}

\begin{document}

\newcommand{\sys}{ILX}
\title{\sys{}: Intelligent "Location+X" Data Systems (Vision Paper)} 

\author{Walid G. Aref$^1$, 
Ahmed M. Aly$^2$, Anas Daghistani$^3$, Yeasir Rayhan$^1$, Jianguo Wang$^1$, Libin Zhou$^1$ }
\affiliation{%
  \institution{$^1$Purdue University, West Lafayette, IN, USA\\
  $^2$Meta Platforms, Menlo Park, CA, USA\\
  $^3$Umm Al-Qura University, Mecca, Saudi Arabia \\
  $^1$\{aref, yrayhan, csjgwang, zhou822\}@purdue.edu,
  $^2$aaly@fb.com,
  $^3$ahdaghistani@uqu.edu.sa
  }
}



\begin{abstract}
\begin{sloppypar}
Due to the ubiquity of mobile phones and location-detection devices, location data is being generated in very large volumes. Queries and operations that are performed on location data warrant the use of database systems.  Despite that, location data is being supported in data systems as an afterthought. 
Typically, relational or NoSQL data
systems that are mostly designed with non-location data in mind get extended with spatial or spatiotemporal indexes, some query operators, and higher level syntactic sugar in order to support location data. The ubiquity of location data and location data services call for systems that are solely designed and optimized for the efficient support of location data.
This paper envisions  designing intelligent location+X data systems, \sys{} for short, where  location is treated as a first-class citizen type. \sys{} is tailored with location data as the main data type (location-first). Because location data is typically augmented with other data types X, e.g., graphs, text data, click streams, annotations, etc., \sys{} needs to be extensible to support other data types X along with location. 
This paper envisions the main features that \sys{} should support, and highlights  research challenges in realizing and supporting \sys{}.
\end{sloppypar}
\end{abstract}

\ccsdesc[500]{Information systems~Database design and models}
\ccsdesc[500]{Information systems~Location based services}
\ccsdesc[300]{Information systems~Geographic information systems}
\ccsdesc[300]{Information systems~Database management system engines}
\ccsdesc[300]{Information systems~Database query processing}
\keywords{machine learning-based location+X system, 
intelligent location servers,
extensible location-based server, 
 query processing, indexing}

\maketitle
\section{Introduction}
\begin{sloppypar}
Location data is ubiquitous due to the popularity of smart phones and location-detection devices. Moreover, location data services are getting into almost all aspects of life, and are getting very sophisticated. This warrants designing data systems that are well-optimized for handling and processing location data, and that  
treat location data as a first-class citizen. Unfortunately, this is not the case. Location data is almost always supported in data systems as an afterthought. Typically, systems that are originally optimized with other objectives in mind eventually get extended to support location data as an afterthought. 
For example, consider the many extensions of relational data systems to support location data. These systems are designed and are optimized to efficiently support relational data with location data being an afterthought add-on feature. Other examples include NoSQL and big data systems, e.g., Hadoop, Spark,  etc., that are designed with other objectives in mind, and with location data either being entirely out of the picture, or is supported as an add-on after the fact. Afterwards, researchers try to fit location data into these systems, and possibly apply some tweaks with sub-optimal extensions to support location data into these systems. Contrast this with designing a system that is mainly optimized from the beginning to support location data as first-class citizen.
A strong analogy would be when starting from a car and tweaking its design to make it fly vs. designing an airplane from scratch, or starting from a helicopter and extending its design to make it function as a submarine in addition to being a helicopter. While these extensions are possible given good engineering and resources, they would not perform as efficient as an airplane or a submarine that are designed from scratch as such.
\end{sloppypar}

\begin{sloppypar}
Not to pick on any other researchers, the first author lists only the systems that his group has developed that follow that same pitfall above, e.g., AQWA~\cite{aqwa-reference} that extends on Hadoop~\cite{hadoop-reference}, LocationSpark~\cite{locationSparkReference} that extends on  Spark~\cite{spark-reference}, Tornado/SWARM~\cite{Tornado-reference, Swarm-reference} that extends over  Apache Storm~\cite{storm-reference}, SP-GiST~\cite{sp-gist-reference} that extends  PostgreSQL~\cite{postgres-reference}, and GRFusion~\cite{GRFusion-EDBT, GRFusion-SIGMOD-demo} that extends over VoltDB~\cite{voltdb-reference}. 
Clearly, Hadoop, Spark, Spark, PostgreSQL, and VoltDB have been originally optimized for non-location data.
Many other researchers and industries follow the same paths with some notable successes, e.g., Oracle Spatial~\cite{Oracle-Spatial}, Spatial Hadoop~\cite{DBLP:conf/icde/EldawyM15}, and GeoSpark~\cite{GeoSpark}.
\end{sloppypar}

\begin{sloppypar}
In recent years, in two keynote talks, the first author of this paper has been advocating for Location+X systems~\cite{GeoRichKeynote2017,MDM2019Keynote} that highlight several research challenges and potential solutions for location-first systems that are to be augmented with other X data types, e.g., graphs, text, and relational data. 
This paper extends beyond the ideas in the two keynotes~\cite{GeoRichKeynote2017, MDM2019Keynote}, and 
presents the vision for \underline{I}ntelligent \underline{L}ocation+\underline{X} systems, \sys{}, for short. The paper highlights the main features of \sys{}, and identifies important research challenges in realizing it.
Notice that there are already existing research that supports \sys{}-like "location-first" vision in some aspects. This paper  helps identify and references these efforts whenever appropriate and as space permits. 
\end{sloppypar}

\section{Highlights of \sys{}}
\sys{} stands for Intelligent Location+X systems. In this section, we highlight each of the components of \sys{}, mainly the Location-first, Intelligence, and Extensibility components, and discuss the features that \sys{} and its surrounding eco-system should support.
\subsection{The "L" in \sys{}: Location}
Location and Location + Time are first-class citizens in \sys{}. In addition to being optimized for operating on location data, \sys{} and its supporting eco-system will provide the following important high-level location-based features.

\subsubsection{\bf 
Protection, Privacy, and the Right to be Forgotten of User's Location Data} 
\begin{sloppypar}
\sys{} should guarantee the privacy of user's location data, e.g., as in~\cite{10.14778/3510397.3510404, 9792226, 10.1145/3474717.3483943}. Users should be able to learn and control what \sys{} knows about them. More generally, \sys{} should support the  General Data Protection Regulation ({\bf GDPR}~\cite{GDPR}).
Privacy and protection of user data and the right to be forgotten should be declaratively and intrinsically  associated with location data in \sys{} and should not be left to the applications to enforce. Lower-level layers of \sys{} should be able to enforce these features, and the user should be able to audit and verify how her/his data is being used and the time it should expire from the system.
\end{sloppypar}

\sys{} should prevent snooping data by implementing an end-to-end encryption technique. Also, \sys{} should not allow users to track the locations of individuals. \sys{} should guarantee to only return query answers that cannot be used to reveal the identity of any user of her/his location. Moreover, an intelligent mechanism should be implemented in \sys{} to detect and block users that are trying to track other users or reveal their identity.


\subsubsection{\bf Discovery,  Integration, and Pricing of Location Data}
\sys{} should be able to support location data lakes~\cite{DBLP:conf/cidr/TerrizzanoSRC15}
   and identify relevant location datasets~\cite{DBLP:conf/cidr/DengFAWSEIMO017,DBLP:conf/icde/FernandezAKYMS18} given user requests. \sys{} should support {\em location data integration} of the discovered datasets.
Query issuers in \sys{} may not worry about which data set to use to answer a certain query. Thus, \sys{}'s  query language should have embedded in it the dataset discovery process. However, what users should be concerned with is the cost of answering their query. 
Location data collection and preparation is costly. Thus, the query execution engine and query optimizer for \sys{} should have cost of data and pricing as an optimization parameter while generating query plans and while discovering and selecting the appropriate location datasets needed to answer location-driven queries.

\subsubsection{\bf Location Data Cleaning and Support for Uncertainty}
Like all other data sources, location data contains many errors and needs cleaning. The \sys{} eco-system should provide cleaning tools for location data, especially the ones uniquely related to location data, e.g., the faulty geographic colocation of two shopping stores in the same location and in the same time duration.
 
In addition to cleaning location data from data entry mistakes, location data is inherently uncertain, e.g., due to accuracy errors in location-detection devices. \sys{}'s query language and execution engine should deal with the uncertainty in location data and  provide query operators that reason given the uncertainty~\cite{DBLP:journals/vldb/SilvaALPA13}.

\begin{sloppypar}
\subsubsection{\bf Location Data Transactional and Online Analytics Support}
\sys{} should offer transactional support due to the heavy update nature of location and related data, e.g., due to objects continuously changing their locations and changing the associated data. Moreover, \sys{} should be able to perform online analytics~\cite{DBLP:journals/tkde/HuangSX04,DBLP:journals/geoinformatica/ShekharLZ03,DBLP:conf/icdm/YooSC05, do-analytical-reference,XieL0LZG16}, especially ones that are unique to location data. 
Finally, because geospatial data is hierarchical in nature (both in space and time), \sys{} should support hierarchical and multi-resolution analytics both in the space and time dimensions.
\end{sloppypar} 

\subsubsection{\bf Human-in-the-loop and Crowdsourcing}
Many transactions in \sys{} will involve human actions~\cite{DBLP:journals/tkde/EltabakhAEO14}. Thus, \sys{} should natively support
Human-in-the-loop, humans-as-query-operators in location-based query evaluation pipelines, and humans as operators in long-standing transactions. \sys{} should protect the location privacy of crowdsourcing workers and tasks, e.g,. as in~\cite{8509301}.

\begin{sloppypar}
\subsubsection{\bf Sampling, Predication, and Approximate Location-data Processing}
Given the massive sizes of location data and demands from location-service for online responses, it may not be feasible to process all location data made available for a given task in a timely fashion. Location-data sampling and approximate query processing techniques should be an integral component of \sys{}'s query execution engine. Tradeoffs between the approximation quality and the runtime requirements of the location service tasks should be well-studied in the context of \sys{}.
\end{sloppypar}

\subsubsection{\bf The Time Dimension, Data Streaming, and Continuous Data Support}
In \sys{}, the time line can be split into three time zones: the past time, the current time (Time NOW), and the future time.
\sys{} should be able to support all three notions of time. 

{\bf Past-time Data}. Past-time data reflects 
historical location data.  \sys{} should be able to store, update, and query historical location data. Workloads in past-time location data are mainly analytical query workloads. 
One good example of this category is the  historical location data that is in the form of moving object trajectories that have taken place in the past. \sys{} should be able to handle these historical location data natively. 

{\bf Current-time Data.}
Current-time data 
is continuously arriving data that reflects what is happening in the time NOW. The workload is heavy in updates to ingest all location data updates that reflect changes of objects' locations over time. 
The workload of current-time data is also heavy in reads in support of continuous queries (that mainly continuously probe current-time data to check current status).
Thus, \sys{} should be able to handle current-time location data workloads that are heavy in both updates and in continuous and snapshot analytics.


{\bf Future Time.} \sys{} should be able to support future 
prediction type of  location data. This is useful for what-if scenarios, decision support, and prediction analytical workloads.

\subsubsection{\bf  3D and 4D Data Support Beyond GeoLocation Data}
\begin{sloppypar}
\sys{} will support spatial data beyond  geolocation data. For example, brain data atlases,   connectivity networks, and brain simulations~\cite{3Dbrain}
are non-geolocation data that fall perfectly within the scope of \sys{}. 
Similarly, geolocation data contains 3D and 4D data, e.g., terrain data and simulations of flood over terrain data. \sys{} dimensionality should extend to support these scenarios.
\end{sloppypar}

\subsubsection{
\bf Visualization}
\sys{} will provide a suite of visualization tools that are tightly-integrated into \sys{}'s query processing and sampling components, e.g., as in~\cite{Kyrix2019,DongBKCL020,GuoFCB18}.
Visualization would support 2D, 3D, and 4D data via animations over time.

\subsection{The "X" in \sys{}: Extensibility}
\begin{sloppypar}
Typically, location data is associated with other data types X, e.g., graphs,   road networks, points of interest,  social network data, click streams, text and tweets, documents, and relational data.
The location engine in \sys{} should be extensible to introduce new data types X as needed by the driving location-service applications. 
Thus, extensibility for adding new data types X will be a first-class feature in \sys{}.
Extensibility will be at all engine levels including storage, indexing, query processing operators, and  query optimization.

\subsubsection{\bf Extensibility in Multi-model Databases}
Recently, multi-model databases have been gaining significant attention in order to address the big varieties in data applications~\cite{MMDBSurvey19}. Example multi-model databases include ArangoDB~\cite{arangodb-reference}, OrientDB~\cite{OrientDB}, BigDAWG~\cite{BigDAWG15}, and Oracle Converged Database~\cite{ConvergedDB}. 

Current multi-model databases either do not support location data at all, or do not support it efficiently, e.g., may support location data as JSON documents, or have geometric data types without indexing support.
Notably, Oracle Converged Database~\cite{ConvergedDB} supports location data with indexing support but is implemented on top of relational tables, which is against the vision and premise of \sys{}.
 

Multi-model database systems support multiple fixed data types within the same system. Having extensibility as a main feature in multi-model databases can be one step in the correct direction.
\end{sloppypar}

\subsubsection{\bf Multi-model Data Stream Support}
Current multi-model data systems do not support data streaming. In addition to being multi-model in nature, e.g., as in~\cite{arangodb-reference}, the multi-model \sys{} should also support both online streaming in addition to the offline processing of location and location+time data. 

\subsection{The "I" in \sys{}: Intelligence}
\begin{sloppypar}
Adopting Machine learning (ML) techniques in systems is a very promising direction given nowadays advances in hardware, GPUs, neural networks, deep learning, and ML software stacks. Location+X systems are no exception. 
Potential benefits for enabling location-first \sys{} systems with ML techniques are multi-fold. 
\end{sloppypar}
\subsubsection{\bf Enhancing over Existing Heuristics}
\begin{sloppypar}
Many location-related  problems involve heuristics that serve as approximations for NP-Complete and NP-Hard problems. Replacing these heuristic solutions with ML-based techniques is expected to produce more efficient and more accurate 
learning-based 
solutions, e.g., as in~\cite{10.1145/3474717.3484217}.
Another example is  handling the dynamic nature and the change in distribution of location data and location queries over time. Reinforcement learning can be used to adapt the underlying organization and partitioning of location data to rebalance the load. One important challenge for using ML in \sys{} is the need for accurate yet real-time  responses to location-based queries. The benefits of augmenting ML into \sys{} in terms of scalability, adaptivity, real-timeliness, and accuracy need to be investigated. 
\end{sloppypar}

\subsubsection{\bf Support for Explainability}
Explainability in AI is an important subject. For the same reasons,  explainability is needed in \sys{} to explain why the ML-based decisions in \sys{} are made, and why other choices are excluded. 
In the broader sense, explainability is needed in \sys{} when choices are made. For example, when a well-known shortest path is not chosen, the user should be given feedback as to why the well-known shortest path has not been chosen, e.g., due to new construction, lane closure, accident, etc.

\subsubsection{\bf Recommendation Operators}
\begin{sloppypar}
\sys{} should support location-driven recommendations and ranking in its query language and execution engine. \sys{}'s query language should have embedded in it personalized recommendation operators, e.g., based on location-aware {\em Collaborative Filtering}, to rank the query engine's responses. 
More generally, recommendations in \sys{} must be aware of the surrounding context that includes not only the locations of objects but also the time of day, the temperature, the dietary restrictions, etc. Means to automatically collect these contexts and means to incorporate user contexts in query processing and in recommendations need to be incorporated into \sys{} to return the most relevant and diversified results to the query issuer, e.g., as in~\cite{10.1007/978-3-319-91563-0_8,Meihui2016,KalamatianosFM21}.
\end{sloppypar}

\section{Infrastructure Highlights of \sys{}}
The massive sizes of location data can flood any location server with data. Thus, one of the main goals in realizing \sys{} is scalability. Scalability in \sys{} will be achieved by a multiplicity of means including adaptivity, elasticity, and the adoption of new hardware and memory platforms, e.g.,  main-memory and persistent memory clusters,  NUMA-awareness,  vectorization, and GPU query processing. \sys{} will adopt  important query processing strategies including federated query processing, query compilation, and  approximate query processing techniques, e.g., distance oracles and other essential location-related query operators. In this section, we briefly highlight these approaches and their roles in \sys{}.

\subsection{Adaptivity, Elasticity, and Memory Disaggregation}
Due to the dynamic changes in location data distributions over space and time, and the occurrence of hot spots, new servers will need to be allocated online while \sys{} is running. Similarly, servers will need to be dynamically deallocated from lightly loaded geospatial regions. Moreover, \sys{} will use disaggregated architectures (Compute servers vs. Memory servers)~\cite{DSM22}, where 
one can add or remove compute or memory independent of each other.

\subsection{Utilizing Modern Hardware}

\subsubsection{\bf NUMA Awareness}
\begin{sloppypar}
Multi-socket systems with non-uniform memory access architectures (NUMA) have been introduced, where each socket is equipped with multiple cores along with its own local memory, and is connected to other sockets, i.e., remote memory with interconnect links. To fully utilize these modern multi-core NUMA hardware, NUMA-aware algorithms~\cite{DBLP:conf/cidr/LiPMRL13,DBLP:conf/sigmod/LeisBK014,DBLP:journals/pvldb/PsaroudakisSMSA16} are continuously being developed for data systems. \sys{} should be designed while considering the characteristics of these multi-core NUMA architectures.
\end{sloppypar}

\subsubsection{\bf Vectorization}
\begin{sloppypar}
Vectorizing a query execution engine~\cite{DBLP:conf/damon/PolychroniouR19} or a standalone database operator, e.g.,  join~\cite{DBLP:journals/pvldb/KimSCKNBLSD09, DBLP:conf/icde/BalkesenTAO13, DBLP:conf/sigmod/BlanasLP11}, scan~\cite{DBLP:journals/pvldb/WillhalmPBPZS09, DBLP:conf/sigmod/ZhouR02}, aggregation~\cite{DBLP:conf/damon/YeRV11, DBLP:conf/sigmod/CieslewiczRSY10, DBLP:conf/damon/PolychroniouR13}, sorting\cite{DBLP:journals/pvldb/ChhuganiNLMHCBKD08, DBLP:conf/sigmod/PolychroniouR14, DBLP:conf/IEEEpact/InoueMKN07, DBLP:conf/sigmod/SatishKCNLKD10, DBLP:journals/pvldb/InoueT15}, Bloom filters~\cite{DBLP:journals/pvldb/LangNKB19} or compression~\cite{DBLP:conf/damon/PolychroniouR15} to utilize data parallelism has gained popularity in recent times due to the introduction of complex SIMD instructions in modern multi-core CPU platforms and the performance gain while executing database queries. VectorWise~\cite{DBLP:conf/cidr/BonczZN05},
DB2 BLU~\cite{DBLP:journals/pvldb/RamanABCKKLLLLMMPSSSSZ13}, columnar SQL Server~\cite{DBLP:conf/sigmod/LarsonCHOPRSZ11},
Quickstep~\cite{DBLP:journals/pvldb/PatelDZPZSMS18} are example data systems that implement vectorization. It is only natural that \sys{}-based systems should benefit from vectorization and its potential can be investigated while designing vectorized location database operators for \sys{}.
\end{sloppypar}


\subsubsection{\bf Main-memory Techniques}
There is a surge of interest in main-memory databases~\cite{MMDB17, Hekaton13, HANA13, PatelDZPZSMS18} because of the dropping price and increasing capacity of main-memory. Thus, it is possible to keep large portions of location data in main-memory for high performance. Many location data techniques need to be revisited for main-memory and the cache hierarchy, e.g., as in~\cite{10.1145/3347146.3359343}.
Caching that is aware of location proximity is called for. Optimizing to minimize CPU cache misses while performing memory-based location operations would be critical to high performance. The main-memory location engine needs to be redesigned because of the lack of need for a buffer manager anymore. Thus, efficient location data layout in main-memory 
and cache-aware location-based indexes
are important factors for a highly performant \sys{} system.

\subsubsection{\bf Persistent Memory}
\begin{sloppypar}
Typically, location-based data systems have been optimized for the traditional memory hierarchy: cache memory, main-memory, disk (or SSD). With the introduction of Persistent and Non-Volatile Memories, e.g., Intel Optane Persistent Memory~\cite{IntelPM}, the traditional memory hierarchy has changed significantly. First, persistent memory is "persistent". Thus, there is no need for disk-like storage. Second, the  speed gap between main-memory and persistent memory is much narrower than what is between main-memory and disk. Thus, location data indexes that have been optimized for disk-based memory hierarchies will need a complete redesign to fit into a persistent-memory-based memory hierarchy, e.g., as in~\cite{LuDLMW21,LuHWL20}. 
Another important factor is that the read and write speeds for persistent memory are not symmetric. Writes are multiple of times slower than Reads.
Location data indexes over persistent memory need to be optimized for that. \end{sloppypar}

\subsubsection{\bf GPUs}
GPUs have a complex memory architecture with various types of memories including texture memory, which is a read-only off-chip memory with caching enabled~\cite{texmem}. Texture cache is specially optimized for 2D spatial locality that makes it an optimal candidate for handling location data in \sys{}.
Designing GPU-friendly data model and algebra by capturing the geometric properties of spatial data to answer spatial queries over large data sets has been gaining popularity~\cite{DBLP:conf/sigmod/DoraiswamyF20} and is in the right direction. \sys{} needs GPU support to naturally make use of the GPU's 2D cache memory that is quite fit for location data. Moreover, GPUs would help subsidize for the expensive geometric data operations, e.g., polygon-polygon intersections, and spatial joins. However, the issue of impedance mismatch between the GPU and CPU memory spaces still need to be addressed to avoid copying data back and forth between the two memory spaces.

\subsection{\bf Query Processing}
In \sys{}, we will adopt several query processing strategies including (a)~{\em Federated Query Processing} due to the variety in the input location data sources, (b)~{\em Multi-model Query Compilation Techniques} to allow location data services and continuous location queries to execute as close to the bare bone of the underlying hardware without multiple software layers that defeat the real-time nature in performing location services, and (c)~{\em Query Operators and Services} unique to the \sys{} environment including Distance Oracles, Map Matching operators, and Address Translation services. Below, we present highlights of these query processing features in \sys{}.

\subsubsection{\bf Federated Query Processing}
\begin{figure}[tbp]
    \centering
    \includegraphics[width=\columnwidth]
    {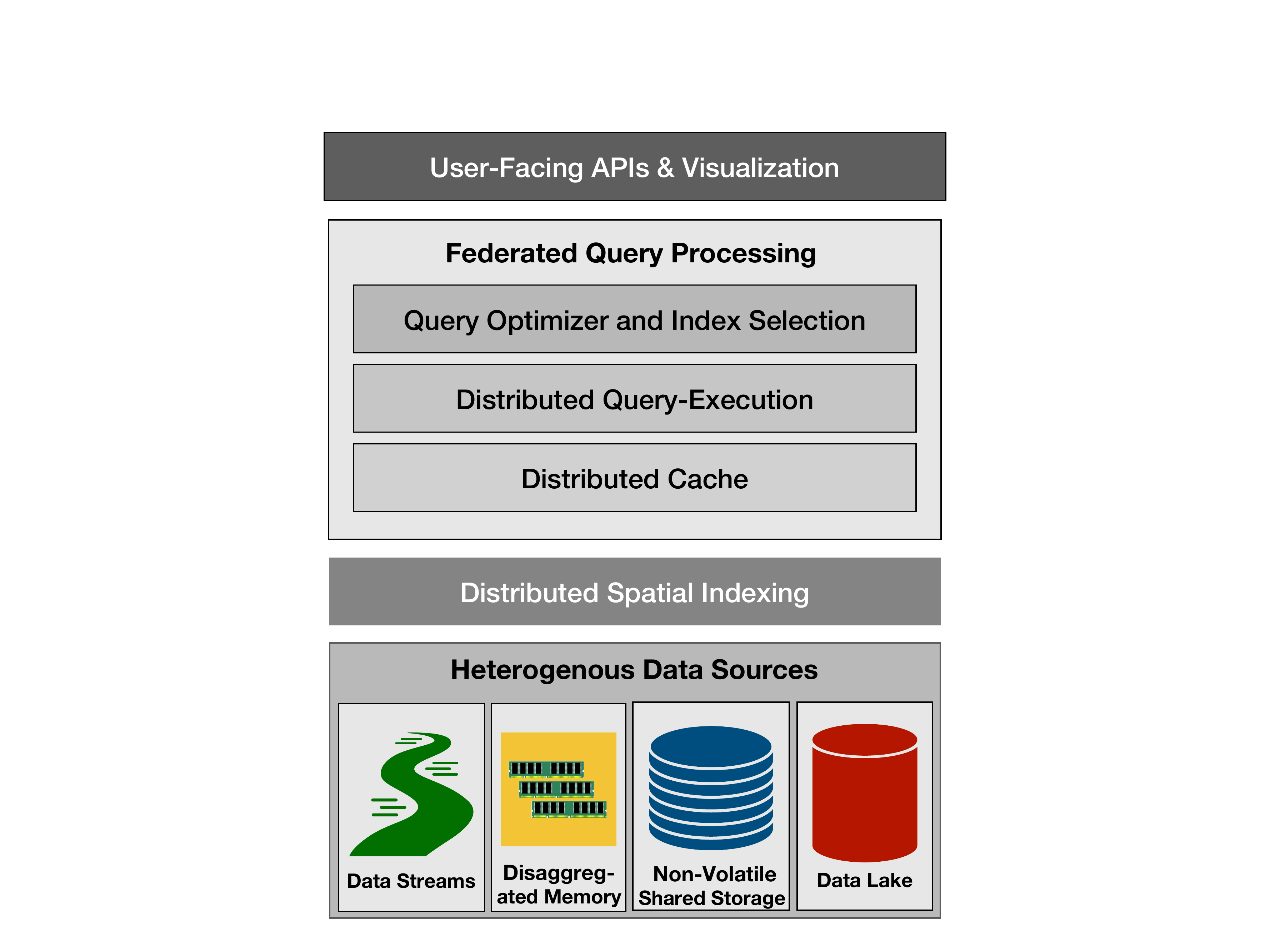}
    \caption{\sys{} federated query processing architecture.}\label{fig:arch}
\end{figure}

\begin{sloppypar}
Federated query processing has been adopted by many recent systems to handle the diversity in the sources of data and be able to process queries across the multiplicity of sources. Example systems are F1~\cite{DBLP:journals/pvldb/ShuteVSHWROLMECRSA13}, Presto~\cite{DBLP:conf/icde/SethiTSPXSYJHSB19}, Flink~\cite{DBLP:journals/debu/CarboneKEMHT15}, and Iceberg~\cite{aiberg}.
The envisioned \sys{} will have a layered architecture that builds on federated query processing. Refer to Figure~\ref{fig:arch}. We briefly explain the stack of layers that constitute \sys{}.
\end{sloppypar}

The user-facing layer offers several APIs for issuing the user queries and receiving the query results. The user-facing layer also offers several visualization tools for presenting the query results using visual representation.  For streaming applications, the query results update the visualization in an online fashion.
The query processing layer consists of three components: a) an optimizer, b) distributed execution, and c) caching. The optimizer is responsible for finding the best plan for the query, as well as finding the best spatial and relational indexes that speed up the execution of the query.  The execution units (i.e., operators) of \sys{} are distributed.  A key feature in \sys{} is that the data is spatially indexed and partitioned according to the spatial features of the data (be it streamed or static), and also according to the query workload distribution.  This partitioning would lead to efficient distributed execution with high throughput and low latency.  Moreover, the caching layer boosts the performance of repeating queries, i.e., these queries that focus on hotspot locations, e.g., downtown areas, event locations, etc., or continuous query evaluation, e.g., in support of data streaming.


\sys{} does not rely on a single data format for its data sources. Thus, \sys{} is founded on a federated query processing platform, where it supports extensible data readers and adapters that can read heterogeneous data formats from different storage and streaming sources.  Moreover, \sys{} can operate on multiple data sources with a variety of formats. A single query can perform a join between a streaming data source and an RDF file from the data lake.

\subsubsection{\bf Multi-model Query Compilation}
\begin{sloppypar}
Query compilation has proven to be quite effective in enhancing the performance of database systems, e.g.,~\cite{Neumann2011, KlonatosKRC14}. In contrast to producing a query evaluation pipeline that the query interpreter executes one-tuple-at-a-time or a vector of tuples at a time, in case vectorization is used, in query compilation, low level C code can be generated and compiled to execute the query. Query compilation eliminates the software interpretation layer, and results in executing the query as close as possible to the bare metal of the hardware.
\end{sloppypar}

There has been good efforts in compiling queries that involve spatial predicates, e.g.,~\cite{TahboubR16}. However,  given the multi-model nature in ILX, query compilation needs to be extended to cover multi-model queries that access , e.g., location, text/JSON, relations, and graph traversal operations.

\subsubsection{\bf 
Query Operators and Services} 
At the core of \sys{} is a set of unique location-related operators that cater to the unique features of location query processing in  \sys{}. These operators include 
{\em Distance Oracle} operators, {\em Map Matching} operators, and {\em Address Translation} services.
We describe each one briefly below. Notice that some of these operators are offered by service providers, e.g., Google Maps GeoLocation APIs~\cite{GMapsGLAPI}. However, they are not open-source, and are provided at a pay-as-you-go pricing model~\cite{GMGLBilling}.

{\bf Distance Oracles.} 
Distance oracles offer a fast means for computing shortest distance in road networks, e.g.,~\cite{do2009-reference, HH05-reference, CH-reference}. Based on the amount of storage allowed for preprocessing, they provide a spectrum of approximate solutions with various error bounds (including 0 error). 
Distance oracles are an integral component for scalability in \sys{}'s query processor. However, current distance oracle technology will need to be extended to allow for operating on arbitrary subsets of the road network, e.g., when a subset of the roads is dynamically selected, e.g., via querying, and then a shortest path computation is required on the selected subset.

{\bf Map Matching Operators.} 
A core operator in \sys{} is the map-matching operator. It maps the physical location of an object to a logical location on the map. Given a road network, say $RN$, and the physical location of an object $O$, e.g., $O$'s longitude, latitude from a GPS reading, say $L_O(long,lat)$, the map matching operator returns from $RN$, the logical location of the object on the map, e.g., the identifier of the road (edge), the intersection (vertex), or the textual address  that $O$ {\em most likely} lies in. 
Notice that there is the possibility of transient errors in the map matching operator due to the inaccuracy in the GPS measurement devices and the misalignment and misregistration of the underlying maps into physical space. Also, the errors depend on whether the map matching operation is performed online or offline. In the case of offline map matching, the entire trajectory of the object is present, and hence it should be more accurate to predict the location of an object at any given point in time. In contrast, in the case of online map matching, only the current and past locations of the object are available for the map matching operator to decide on the logical location on the map of an object at current time. Hence, in the online case, the map matching operator is prone to more errors. 
The map matching operator is commonly used in GPS devices to display the location of the object on the logical map and to help with the vehicle navigation process using the logical map as a guide. It is anticipated that \sys{} will also make heavy use of this operator at both the  query processing and optimization levels. Many useful map matching operators exist that we plan to utilize and build on from within \sys{}, e.g.,~\cite{NewsonK09, YuanZZXS10, OsogamiR13, LouZZXWH09, BrakatsoulasPSW05}. 

{\bf Address Translation Operators.} Another important and useful building block for query processing and optimization in \sys{} is the address translation operator. This operator is the inverse of the map matching operator. Given a textual address input, this operator returns the address's corresponding longitude and latitude.

The distance oracle, map matching, and address translation operations will be used extensively in query processing within \sys{}.

\subsection{Location-based Access Methods}
\subsubsection{\bf Clustered Location Data Indexes}
Access methods and indexes for location data are essential components in~\sys{}. However, with the location data type  being a first-class citizen in \sys{}, location data indexes need to become the primary storage methods and clustered indexes that 
host all the other types of data in addition to the location data. For example, if \sys{} has a quad-tree index to store the coordinates for a point data set, the same quad-tree could serve as a clustered index that also stores the entire description of the point data objects, e.g., the city names, the city population, etc., inside the index. Additional indexing methods will be based on the types X associated with the location data. However, clustering of data will be location-driven.

\subsubsection{\bf Update-Intensive Indexing Techniques} 
\begin{sloppypar}
The continuous move and change in location of objects in space over time results in an update-intensive workload.
Thus, an important feature in location access methods is the support for update-intensive  indexing. 
Techniques exist for handling frequent updates in location indexes, e.g.,~\cite{DBLP:journals/vldb/SilvaXA09}. However, they need to be extended to support (1)~memory-based location indexes, (2)~become cache- and NUMA-aware, and (3)~be optimized for disaggregated memory. Disaggregation has become feasible and practical due to the successful use of high-speed remote direct memory access (RDMA) over the network~\cite{DSM22}. Location data indexes need to be adapted to support the  disaggregated architecture over RDMA.
\end{sloppypar}

\subsubsection{\bf LSM-based Location Indexes} LSM indexes~\cite{DBLP:journals/acta/ONeilCGO96} are optimized for write-intensive key-value workloads. Because location serves as a secondary key, to be effective, LSM indexes need to be adapted in support of update-intensive secondary-key location-data workloads~\cite{Shin0A21}. 

\subsubsection{\bf Location-based Learned Indexes}
\begin{sloppypar}
Machine Learning (ML) techniques have been applied successfully to build various types of learned indexes~\cite{kraska2018case}.
It has been extended to the multi-dimensional case, e.g.,~\cite{al2020tutorial,Abdullah2022}. Learned indexes have shown potential in terms of smaller index size and faster performance in contrast to traditional indexes.
Learned indexes work well for static data sets as training of the learned models take place in a preprocessing phase.
Realizing learned indexes for dynamic data sets has been a challenge due to the need to continuously retrain the models.
There are some very successful attempts to deal with dynamic data in the multidimensional case. Of mention are LISA~\cite{Li0ZY020} and RSMI~\cite{qi13effectively}. \sys{} needs to adopt similar ideas, and extend 
these learned indexes to accommodate the time dimension to be able to handle real-time trajectory data. 
\end{sloppypar}

\subsection{Concurrency Control, Integrity, and Fault Tolerance}
Concurrency control plays a critical role in \sys{} to coordinate concurrent read and write operations for scalability. Many existing concurrency control protocols in spatial databases are based on locking data objects (e.g.,~\cite{DaiL17, ChaudhryY22, ChakrabartiM99, SongKY04}). 
Two possible approaches can be explored in \sys{}. First, in contrast to locking data objects, \sys{} can consider locking the underlying physical space or specific locations in space under the premise that no two objects can share the underlying physical space at the same time. In contrast to data-driven locking, locking physical locations can resemble {\em space-driven locking} in location data indexes that have disjoint space-driven partitioning of the underlying space. 
Two issues remain to be addressed for this approach to be credible: 
(1)~Handle consistently the issue of multi-granularity locking in the physical space and (2)~Handle the issue of location uncertainty. If the location of an object is uncertain or is not measured precisely, then locking of physical locations may not have one-to-one correspondence with the locations of the objects as stored within \sys{} or within \sys{}'s location data indexes. This may introduce overlaps in potential locations of where  objects might be in space.  
More research is needed to address the issue of uncertainty in conjunction with physical location locking and the location overlaps it introduces.

The second approach that needs to be explored in \sys{} is to adopt concurrency control techniques that can scale to  hundreds and thousands of cores~\cite{BangMPB20, YuBPDS14}. It is important to design lock-free concurrency control for spatial access methods along the same lines as the lock-free B-tree (the Bw-tree~\cite{LevandoskiLS13a}).

Finally, the new infrastructure that \sys{} will be deployed in poses additional challenges for concurrency control. For example, in the RDMA-enabled disaggregated memory architecture~\cite{CaoZYLWHCCLFWWS21, DSM22}, it is non-trivial to lock the remote objects using RDMA primitives, and hence existing concurrency control protocols need to be revisited.

\sys{} should be able to tolerate faults, e.g., via replication.   \sys{} should be able to recover its indexes if they get partially or completely lost or damaged due upon faulting. Recovering from faults are to be performed online without system shutdown and while guaranteeing correctness of the system operation e.g., during online repartitioning of data, \sys{} should guarantee that no data gets lost and no data is reported twice as part of an answer to a query.

\subsection{Location Data Compression}
Data compression is an important technique especially for spatial databases due to the huge amount of location data. It not only can save memory but also can improve query time due to the smaller data sizes being retrieved. Compression is highly under-studied in spatial databases~\cite{Lin4568473, Chovanec5678116}. It requires a systematic study of compression techniques for both location data and location data indexes. Although there are some compression algorithms for floating-point data~\cite{Chimp22, PelkonenFCHMTV15}, it is not clear how they perform on location data because these algorithms usually work well on specific data distributions. For location indexes, e.g., the R-tree, it is important to compress the structural information, similar to B-tree structural compression~\cite{BayerU77, Lomet01}. Another important design consideration is to support query processing on compressed data and indexes, which will improve the performance. More research is needed to evaluate the impact to compression ratio.

\subsection{Semantics and RDF-based Location Data}
\begin{sloppypar}
Many geospatial datasets are part of the Web of Data. Several geospatial extensions to the SPARQL query language have been introduced to query and reason over geospatial semantic data. \sys{} should be able to natively store and reason over geospatial RDF data. It is important for \sys{} to handle the slight geo-semantic inaccuracies, e.g., the predicate "north-of" can roughly describe objects that are slightly towards the northeast direction. \sys{} should be able to reason over location data given these semantic ambiguities. Moreover, \sys{} should be able to make use of the  interlinked  topological relations in the Linked Open Data cloud (LOD), and help produce new  geospatial interlinks progressively in LOD as a side effect, e.g., as in~\cite{10.1145/3510025}. 
\end{sloppypar}


\subsection{Security and Resilience to Attacks}
\sys{} should be resilient to malicious activities, e.g., attacks to stop the system, alter, or snoop data. Systems that use dynamic load balancing mechanisms are vulnerable to malicious attacks~\cite{daghistani2021guard}. This type of attack affects system availability. Attackers can make the system in continuous state of rebalancing. Other types of attacks that can affect \sys{} need to be investigated, e.g., faking the location of data, hiding the detection of important location data by flooding the system with irrelevant data in the same location. \sys{} should be resilient to these attacks by having intelligence to detect and block malicious users. It should analyze user behavior as individuals and as groups to detect and prevent any malicious activities. 

\subsection{Useful EcoSystem Tools}
Various geometrical and spatiotemporal toolkits and libraries exist, e.g.,~\cite{H3,S2,LocationTech}, that can be partly useful for the \sys{} ecosystem. Also, location data generators, e.g.,~\cite{10.1007/978-3-642-40235-7_3}, would be an integral part of the \sys{} ecosystem.

\section{Summary} 
This paper highlights the main features and challenges in realizing \sys{}-like systems. Several existing research works follow some aspects of the \sys{} vision, and hence are in the right direction. Due to space limitation, not all of these research works are cited in this paper. However, this paper helps identify such works. 

Benchmarks for testing and tuning the performance of all of \sys{}'s features will be an integral part of  \sys{}'s ecosystem. Many such benchmarks already exist in the literature. However, once \sys{} is realized, targeted micro-benchmarks for specific features of \sys{} will need to be developed.

\section{Acknowledgements}
The authors acknowledge the support of the U.S. National Science Foundation under Grant Numbers III-1815796 and IIS-1910216.

\clearpage\newpage
\balance

\begin{sloppypar}
\bibliographystyle{ACM-Reference-Format}
\bibliography{authordraft}


\begin{thebibliography}{124}


\ifx \showCODEN    \undefined \def \showCODEN     #1{\unskip}     \fi
\ifx \showDOI      \undefined \def \showDOI       #1{#1}\fi
\ifx \showISBNx    \undefined \def \showISBNx     #1{\unskip}     \fi
\ifx \showISBNxiii \undefined \def \showISBNxiii  #1{\unskip}     \fi
\ifx \showISSN     \undefined \def \showISSN      #1{\unskip}     \fi
\ifx \showLCCN     \undefined \def \showLCCN      #1{\unskip}     \fi
\ifx \shownote     \undefined \def \shownote      #1{#1}          \fi
\ifx \showarticletitle \undefined \def \showarticletitle #1{#1}   \fi
\ifx \showURL      \undefined \def \showURL       {\relax}        \fi
\providecommand\bibfield[2]{#2}
\providecommand\bibinfo[2]{#2}
\providecommand\natexlab[1]{#1}
\providecommand\showeprint[2][]{arXiv:#2}

\bibitem[\protect\citeauthoryear{??}{H3}{[n.d.]}]%
        {H3}
 \bibinfo{year}{[n.d.]}\natexlab{}.
\newblock \bibinfo{title}{H3: Uber’s Hexagonal Hierarchical Spatial Index}.
\newblock \bibinfo{howpublished}{\url{https://eng.uber.com/}\\ \url{h3/}}.
\newblock


\bibitem[\protect\citeauthoryear{??}{Loc}{[n.d.]}]%
        {LocationTech}
 \bibinfo{year}{[n.d.]}\natexlab{}.
\newblock \bibinfo{title}{The LocationTech Technology (LTT)}.
\newblock \bibinfo{howpublished}{\url{https://projects.eclipse.org/projec}\\
  \url{ts/locationtech}}.
\newblock


\bibitem[\protect\citeauthoryear{??}{S2}{[n.d.]}]%
        {S2}
 \bibinfo{year}{[n.d.]}\natexlab{}.
\newblock \bibinfo{title}{S2 Geometry}.
\newblock \bibinfo{howpublished}{\url{https://s2geometry.io}}.
\newblock


\bibitem[\protect\citeauthoryear{??}{Con}{2020}]%
        {ConvergedDB}
 \bibinfo{year}{2020}\natexlab{}.
\newblock \bibinfo{title}{{Oracle's Converged Database: How to Make Developers
  And Data More Productive.
  \url{https://www.oracle.com/a/otn/docs/databas/oracle-converge}
  \\\url{database-technicalbrief.pdf}}}.
\newblock
\newblock


\bibitem[\protect\citeauthoryear{??}{aib}{2022}]%
        {aiberg}
 \bibinfo{year}{2022}\natexlab{}.
\newblock \bibinfo{title}{Apache Iceberg}.
\newblock
  \bibinfo{howpublished}{\url{https://www.dremio.com/resources/guides/apache-ice}\\{\url{berg-an-architectural-look-under-the-covers/}}}.
\newblock


\bibitem[\protect\citeauthoryear{??}{sto}{2022}]%
        {storm-reference}
 \bibinfo{year}{2022}\natexlab{}.
\newblock \bibinfo{title}{Apache Storm}.
\newblock \bibinfo{howpublished}{\url{https://storm.apache.org/}}.
\newblock


\bibitem[\protect\citeauthoryear{??}{ara}{2022}]%
        {arangodb-reference}
 \bibinfo{year}{2022}\natexlab{}.
\newblock \bibinfo{title}{ArangoDB}.
\newblock \bibinfo{howpublished}{\url{https://www.arangodb.com/}}.
\newblock


\bibitem[\protect\citeauthoryear{??}{GDP}{2022}]%
        {GDPR}
 \bibinfo{year}{2022}\natexlab{}.
\newblock \bibinfo{title}{General Data Protection Regulation (GDPR)}.
\newblock \bibinfo{howpublished}{\url{https://gdpr-info.eu/art-1-g} \\
  \url{dpr}}.
\newblock


\bibitem[\protect\citeauthoryear{??}{GMa}{2022}]%
        {GMapsGLAPI}
 \bibinfo{year}{2022}\natexlab{}.
\newblock \bibinfo{title}{GeoLocation API}.
\newblock
  \bibinfo{howpublished}{\url{https://developers.google.com/maps/documentation/ge}\\
  \url{olocation/overview}}.
\newblock


\bibitem[\protect\citeauthoryear{??}{GMG}{2022}]%
        {GMGLBilling}
 \bibinfo{year}{2022}\natexlab{}.
\newblock \bibinfo{title}{GeoLocation API: Usage and Billing}.
\newblock \bibinfo{howpublished}{\url{https://developers.google.com/map}\\
  \url{s/documentation/geolocation/usage-and-billing}}.
\newblock


\bibitem[\protect\citeauthoryear{??}{Int}{2022}]%
        {IntelPM}
 \bibinfo{year}{2022}\natexlab{}.
\newblock \bibinfo{title}{Intel Optane Persistent Memory}.
\newblock \bibinfo{howpublished}{\url{https://www.intel.com/content/www/u}
  \\\url{s/en/architecture-and-technology/optane-dc-persistent-memory.html}}.
\newblock


\bibitem[\protect\citeauthoryear{??}{Ora}{2022}]%
        {Oracle-Spatial}
 \bibinfo{year}{2022}\natexlab{}.
\newblock \bibinfo{title}{Oracle Spatial}.
\newblock
  \bibinfo{howpublished}{\url{https://www.oracle.com/database/spatial/}}.
\newblock


\bibitem[\protect\citeauthoryear{??}{Ori}{2022}]%
        {OrientDB}
 \bibinfo{year}{2022}\natexlab{}.
\newblock \bibinfo{title}{OrientDB}.
\newblock \bibinfo{howpublished}{\url{https://orientdb.org/}}.
\newblock


\bibitem[\protect\citeauthoryear{??}{tex}{2022}]%
        {texmem}
 \bibinfo{year}{2022}\natexlab{}.
\newblock \bibinfo{title}{Texture Memory in GPU}.
\newblock
  \bibinfo{howpublished}{\url{https://docs.nvidia.com/gameworks/content/de}\\
  \url{velopertools/desktop/analysis/report/cudaexperiments/kernellevel/memoryst}\\
  \url{atisticstexture.htm}}.
\newblock


\bibitem[\protect\citeauthoryear{Al{-}Mamun, Haider, Wang, and Aref}{Al{-}Mamun
  et~al\mbox{.}}{2022}]%
        {Abdullah2022}
\bibfield{author}{\bibinfo{person}{Abdullah Al{-}Mamun}, \bibinfo{person}{Ch.
  Md.~Rakin Haider}, \bibinfo{person}{Jianguo Wang}, {and}
  \bibinfo{person}{Walid~G. Aref}.} \bibinfo{year}{2022}\natexlab{}.
\newblock \showarticletitle{{The "AI+R"-tree: An Instance-optimized R-tree}}.
  In \bibinfo{booktitle}{\emph{MDM}}.
\newblock


\bibitem[\protect\citeauthoryear{Al-Mamun, Wu, and Aref}{Al-Mamun
  et~al\mbox{.}}{2020}]%
        {al2020tutorial}
\bibfield{author}{\bibinfo{person}{Abdullah Al-Mamun}, \bibinfo{person}{Hao
  Wu}, {and} \bibinfo{person}{Walid~G Aref}.} \bibinfo{year}{2020}\natexlab{}.
\newblock \showarticletitle{A tutorial on learned multi-dimensional indexes}.
  In \bibinfo{booktitle}{\emph{{SIGSPATIAL}}}. \bibinfo{pages}{1--4}.
\newblock


\bibitem[\protect\citeauthoryear{Aly, Mahmood, Hassan, Aref, Ouzzani,
  Elmeleegy, and Qadah}{Aly et~al\mbox{.}}{2015}]%
        {aqwa-reference}
\bibfield{author}{\bibinfo{person}{Ahmed~M. Aly}, \bibinfo{person}{Ahmed~R.
  Mahmood}, \bibinfo{person}{Mohamed~S. Hassan}, \bibinfo{person}{Walid~G.
  Aref}, \bibinfo{person}{Mourad Ouzzani}, \bibinfo{person}{Hazem Elmeleegy},
  {and} \bibinfo{person}{Thamir Qadah}.} \bibinfo{year}{2015}\natexlab{}.
\newblock \showarticletitle{{AQWA:} Adaptive Query-Workload-Aware Partitioning
  of Big Spatial Data}.
\newblock \bibinfo{journal}{\emph{{PVLDB}}} \bibinfo{volume}{8},
  \bibinfo{number}{13} (\bibinfo{year}{2015}), \bibinfo{pages}{2062--2073}.
\newblock


\bibitem[\protect\citeauthoryear{Aref}{Aref}{2017}]%
        {GeoRichKeynote2017}
\bibfield{author}{\bibinfo{person}{Walid~G. Aref}.}
  \bibinfo{year}{2017}\natexlab{}.
\newblock \showarticletitle{Location + X Big Data Systems: Challenges and Some
  Solutions}. In \bibinfo{booktitle}{\emph{{GeoRich}}}.
\newblock


\bibitem[\protect\citeauthoryear{Aref}{Aref}{2019}]%
        {MDM2019Keynote}
\bibfield{author}{\bibinfo{person}{Walid~G. Aref}.}
  \bibinfo{year}{2019}\natexlab{}.
\newblock \showarticletitle{The Dos and Don’ts of Spatial+X Data Management:
  A “Systems Perspective”}. In \bibinfo{booktitle}{\emph{{MDM}}}.
\newblock


\bibitem[\protect\citeauthoryear{Aref and Ilyas}{Aref and Ilyas}{2001}]%
        {sp-gist-reference}
\bibfield{author}{\bibinfo{person}{Walid~G. Aref} {and}
  \bibinfo{person}{Ihab~F. Ilyas}.} \bibinfo{year}{2001}\natexlab{}.
\newblock \showarticletitle{SP-GiST: An Extensible Database Index for
  Supporting Space Partitioning Trees}.
\newblock \bibinfo{journal}{\emph{J. Intell. Inf. Syst.}} \bibinfo{volume}{17},
  \bibinfo{number}{2-3} (\bibinfo{year}{2001}), \bibinfo{pages}{215--240}.
\newblock


\bibitem[\protect\citeauthoryear{Balkesen, Teubner, Alonso, and
  {\"{O}}zsu}{Balkesen et~al\mbox{.}}{2013}]%
        {DBLP:conf/icde/BalkesenTAO13}
\bibfield{author}{\bibinfo{person}{Cagri Balkesen}, \bibinfo{person}{Jens
  Teubner}, \bibinfo{person}{Gustavo Alonso}, {and} \bibinfo{person}{M.~Tamer
  {\"{O}}zsu}.} \bibinfo{year}{2013}\natexlab{}.
\newblock \showarticletitle{Main-memory hash joins on multi-core CPUs: Tuning
  to the underlying hardware}. In \bibinfo{booktitle}{\emph{{ICDE}}}.
  \bibinfo{pages}{362--373}.
\newblock


\bibitem[\protect\citeauthoryear{Bang, May, Petrov, and Binnig}{Bang
  et~al\mbox{.}}{2020}]%
        {BangMPB20}
\bibfield{author}{\bibinfo{person}{Tiemo Bang}, \bibinfo{person}{Norman May},
  \bibinfo{person}{Ilia Petrov}, {and} \bibinfo{person}{Carsten Binnig}.}
  \bibinfo{year}{2020}\natexlab{}.
\newblock \showarticletitle{The tale of 1000 Cores: an evaluation of
  concurrency control on real(ly) large multi-socket hardware}. In
  \bibinfo{booktitle}{\emph{{DaMoN}}}. \bibinfo{pages}{3:1--3:9}.
\newblock


\bibitem[\protect\citeauthoryear{Bayer and Unterauer}{Bayer and
  Unterauer}{1977}]%
        {BayerU77}
\bibfield{author}{\bibinfo{person}{Rudolf Bayer} {and} \bibinfo{person}{Karl
  Unterauer}.} \bibinfo{year}{1977}\natexlab{}.
\newblock \showarticletitle{Prefix B-Trees}.
\newblock \bibinfo{journal}{\emph{{TODS}}} \bibinfo{volume}{2},
  \bibinfo{number}{1} (\bibinfo{year}{1977}), \bibinfo{pages}{11--26}.
\newblock


\bibitem[\protect\citeauthoryear{Blanas, Li, and Patel}{Blanas
  et~al\mbox{.}}{2011}]%
        {DBLP:conf/sigmod/BlanasLP11}
\bibfield{author}{\bibinfo{person}{Spyros Blanas}, \bibinfo{person}{Yinan Li},
  {and} \bibinfo{person}{Jignesh~M. Patel}.} \bibinfo{year}{2011}\natexlab{}.
\newblock \showarticletitle{Design and evaluation of main memory hash join
  algorithms for multi-core CPUs}. In \bibinfo{booktitle}{\emph{{SIGMOD}}}.
  \bibinfo{pages}{37--48}.
\newblock


\bibitem[\protect\citeauthoryear{Boncz, Zukowski, and Nes}{Boncz
  et~al\mbox{.}}{2005}]%
        {DBLP:conf/cidr/BonczZN05}
\bibfield{author}{\bibinfo{person}{Peter~A. Boncz}, \bibinfo{person}{Marcin
  Zukowski}, {and} \bibinfo{person}{Niels Nes}.}
  \bibinfo{year}{2005}\natexlab{}.
\newblock \showarticletitle{MonetDB/X100: Hyper-Pipelining Query Execution}. In
  \bibinfo{booktitle}{\emph{{CIDR}}}. \bibinfo{pages}{225--237}.
\newblock


\bibitem[\protect\citeauthoryear{Brakatsoulas, Pfoser, Salas, and
  Wenk}{Brakatsoulas et~al\mbox{.}}{2005}]%
        {BrakatsoulasPSW05}
\bibfield{author}{\bibinfo{person}{Sotiris Brakatsoulas},
  \bibinfo{person}{Dieter Pfoser}, \bibinfo{person}{Randall Salas}, {and}
  \bibinfo{person}{Carola Wenk}.} \bibinfo{year}{2005}\natexlab{}.
\newblock \showarticletitle{On Map-Matching Vehicle Tracking Data}. In
  \bibinfo{booktitle}{\emph{{PVLDB}}}. \bibinfo{pages}{853--864}.
\newblock


\bibitem[\protect\citeauthoryear{Cao, Zhang, Yang, Li, Wang, Hu, Cheng, Chen,
  Liu, Fang, Wang, Wang, Sun, Yang, Cheng, Chen, Wu, Hu, Zhao, Gao, Cai, Zhang,
  and Tong}{Cao et~al\mbox{.}}{2021}]%
        {CaoZYLWHCCLFWWS21}
\bibfield{author}{\bibinfo{person}{Wei Cao}, \bibinfo{person}{Yingqiang Zhang},
  \bibinfo{person}{Xinjun Yang}, \bibinfo{person}{Feifei Li},
  \bibinfo{person}{Sheng Wang}, \bibinfo{person}{Qingda Hu},
  \bibinfo{person}{Xuntao Cheng}, \bibinfo{person}{Zongzhi Chen},
  \bibinfo{person}{Zhenjun Liu}, \bibinfo{person}{Jing Fang},
  \bibinfo{person}{Bo Wang}, \bibinfo{person}{Yuhui Wang},
  \bibinfo{person}{Haiqing Sun}, \bibinfo{person}{Ze Yang},
  \bibinfo{person}{Zhushi Cheng}, \bibinfo{person}{Sen Chen},
  \bibinfo{person}{Jian Wu}, \bibinfo{person}{Wei Hu}, \bibinfo{person}{Jianwei
  Zhao}, \bibinfo{person}{Yusong Gao}, \bibinfo{person}{Songlu Cai},
  \bibinfo{person}{Yunyang Zhang}, {and} \bibinfo{person}{Jiawang Tong}.}
  \bibinfo{year}{2021}\natexlab{}.
\newblock \showarticletitle{PolarDB Serverless: {A} Cloud Native Database for
  Disaggregated Data Centers}. In \bibinfo{booktitle}{\emph{{SIGMOD}}}.
  \bibinfo{pages}{2477--2489}.
\newblock


\bibitem[\protect\citeauthoryear{Carbone, Katsifodimos, Ewen, Markl, Haridi,
  and Tzoumas}{Carbone et~al\mbox{.}}{2015}]%
        {DBLP:journals/debu/CarboneKEMHT15}
\bibfield{author}{\bibinfo{person}{Paris Carbone}, \bibinfo{person}{Asterios
  Katsifodimos}, \bibinfo{person}{Stephan Ewen}, \bibinfo{person}{Volker
  Markl}, \bibinfo{person}{Seif Haridi}, {and} \bibinfo{person}{Kostas
  Tzoumas}.} \bibinfo{year}{2015}\natexlab{}.
\newblock \showarticletitle{Apache Flink{\texttrademark}: Stream and Batch
  Processing in a Single Engine}.
\newblock \bibinfo{journal}{\emph{{IEEE} Data Eng. Bull.}}
  \bibinfo{volume}{38}, \bibinfo{number}{4} (\bibinfo{year}{2015}),
  \bibinfo{pages}{28--38}.
\newblock


\bibitem[\protect\citeauthoryear{Chakrabarti and Mehrotra}{Chakrabarti and
  Mehrotra}{1999}]%
        {ChakrabartiM99}
\bibfield{author}{\bibinfo{person}{Kaushik Chakrabarti} {and}
  \bibinfo{person}{Sharad Mehrotra}.} \bibinfo{year}{1999}\natexlab{}.
\newblock \showarticletitle{Efficient Concurrency Control in Multidimensional
  Access Methods}. In \bibinfo{booktitle}{\emph{{SIGMOD}}}.
  \bibinfo{pages}{25--36}.
\newblock


\bibitem[\protect\citeauthoryear{Chaudhry and Yousaf}{Chaudhry and
  Yousaf}{2022}]%
        {ChaudhryY22}
\bibfield{author}{\bibinfo{person}{Natalia Chaudhry} {and}
  \bibinfo{person}{Muhammad~Murtaza Yousaf}.} \bibinfo{year}{2022}\natexlab{}.
\newblock \showarticletitle{Concurrency control for real-time and mobile
  transactions: Historical view, challenges, and evolution of practices}.
\newblock \bibinfo{journal}{\emph{Concurr. Comput. Pract. Exp.}}
  \bibinfo{volume}{34}, \bibinfo{number}{3} (\bibinfo{year}{2022}).
\newblock


\bibitem[\protect\citeauthoryear{Chhugani, Nguyen, Lee, Macy, Hagog, Chen,
  Baransi, Kumar, and Dubey}{Chhugani et~al\mbox{.}}{2008}]%
        {DBLP:journals/pvldb/ChhuganiNLMHCBKD08}
\bibfield{author}{\bibinfo{person}{Jatin Chhugani}, \bibinfo{person}{Anthony~D.
  Nguyen}, \bibinfo{person}{Victor~W. Lee}, \bibinfo{person}{William Macy},
  \bibinfo{person}{Mostafa Hagog}, \bibinfo{person}{Yen{-}Kuang Chen},
  \bibinfo{person}{Akram Baransi}, \bibinfo{person}{Sanjeev Kumar}, {and}
  \bibinfo{person}{Pradeep Dubey}.} \bibinfo{year}{2008}\natexlab{}.
\newblock \showarticletitle{Efficient implementation of sorting on multi-core
  {SIMD} {CPU} architecture}.
\newblock \bibinfo{journal}{\emph{{PVLDB}}} \bibinfo{volume}{1},
  \bibinfo{number}{2} (\bibinfo{year}{2008}), \bibinfo{pages}{1313--1324}.
\newblock


\bibitem[\protect\citeauthoryear{Chovanec, Krátký, and Walder}{Chovanec
  et~al\mbox{.}}{2010}]%
        {Chovanec5678116}
\bibfield{author}{\bibinfo{person}{Peter Chovanec}, \bibinfo{person}{Michal
  Krátký}, {and} \bibinfo{person}{Jiří Walder}.}
  \bibinfo{year}{2010}\natexlab{}.
\newblock \showarticletitle{Lossless R-tree compression using variable-length
  codes}. In \bibinfo{booktitle}{\emph{{ICITST}}}. \bibinfo{pages}{1--8}.
\newblock


\bibitem[\protect\citeauthoryear{Cieslewicz, Ross, Satsumi, and Ye}{Cieslewicz
  et~al\mbox{.}}{2010}]%
        {DBLP:conf/sigmod/CieslewiczRSY10}
\bibfield{author}{\bibinfo{person}{John Cieslewicz},
  \bibinfo{person}{Kenneth~A. Ross}, \bibinfo{person}{Kyoho Satsumi}, {and}
  \bibinfo{person}{Yang Ye}.} \bibinfo{year}{2010}\natexlab{}.
\newblock \showarticletitle{Automatic contention detection and amelioration for
  data-intensive operations}. In \bibinfo{booktitle}{\emph{{SIGMOD}}}.
  \bibinfo{pages}{483--494}.
\newblock


\bibitem[\protect\citeauthoryear{Daghistani, Aref, Ghafoor, and
  Mahmood}{Daghistani et~al\mbox{.}}{2021a}]%
        {Swarm-reference}
\bibfield{author}{\bibinfo{person}{Anas Daghistani}, \bibinfo{person}{Walid~G.
  Aref}, \bibinfo{person}{Arif Ghafoor}, {and} \bibinfo{person}{Ahmed~R.
  Mahmood}.} \bibinfo{year}{2021}\natexlab{a}.
\newblock \showarticletitle{{SWARM:} Adaptive Load Balancing in Distributed
  Streaming Systems for Big Spatial Data}.
\newblock \bibinfo{journal}{\emph{{ACM} TSAS}} \bibinfo{volume}{7},
  \bibinfo{number}{3} (\bibinfo{year}{2021}), \bibinfo{pages}{14:1--14:43}.
\newblock


\bibitem[\protect\citeauthoryear{Daghistani, Khayat, Felemban, Aref, and
  Ghafoor}{Daghistani et~al\mbox{.}}{2021b}]%
        {daghistani2021guard}
\bibfield{author}{\bibinfo{person}{Anas Daghistani}, \bibinfo{person}{Mosab
  Khayat}, \bibinfo{person}{Muhamad Felemban}, \bibinfo{person}{Walid~G Aref},
  {and} \bibinfo{person}{Arif Ghafoor}.} \bibinfo{year}{2021}\natexlab{b}.
\newblock \showarticletitle{Guard: Attack-Resilient Adaptive Load Balancing in
  Distributed Streaming Systems}.
\newblock \bibinfo{journal}{\emph{IEEE TDSC}} (\bibinfo{year}{2021}).
\newblock


\bibitem[\protect\citeauthoryear{Dai and Lu}{Dai and Lu}{2017}]%
        {DaiL17}
\bibfield{author}{\bibinfo{person}{Jing~(David) Dai} {and}
  \bibinfo{person}{Chang{-}Tien Lu}.} \bibinfo{year}{2017}\natexlab{}.
\newblock \showarticletitle{Concurrency Control for Spatial Access}.
\newblock In \bibinfo{booktitle}{\emph{Encyclopedia of {GIS}}}.
  \bibinfo{pages}{285--286}.
\newblock


\bibitem[\protect\citeauthoryear{Daniel, Matera, Quintarelli, Tanca, and
  Zaccaria}{Daniel et~al\mbox{.}}{2018}]%
        {10.1007/978-3-319-91563-0_8}
\bibfield{author}{\bibinfo{person}{Florian Daniel}, \bibinfo{person}{Maristella
  Matera}, \bibinfo{person}{Elisa Quintarelli}, \bibinfo{person}{Letizia
  Tanca}, {and} \bibinfo{person}{Vittorio Zaccaria}.}
  \bibinfo{year}{2018}\natexlab{}.
\newblock \showarticletitle{Context-Aware Access to Heterogeneous Resources
  Through On-the-Fly Mashups}. In \bibinfo{booktitle}{\emph{CAiSE}}.
  \bibinfo{pages}{119--134}.
\newblock


\bibitem[\protect\citeauthoryear{Deng, Fernandez, Abedjan, Wang, Stonebraker,
  Elmagarmid, Ilyas, Madden, Ouzzani, and Tang}{Deng et~al\mbox{.}}{2017}]%
        {DBLP:conf/cidr/DengFAWSEIMO017}
\bibfield{author}{\bibinfo{person}{Dong Deng}, \bibinfo{person}{Raul~Castro
  Fernandez}, \bibinfo{person}{Ziawasch Abedjan}, \bibinfo{person}{Sibo Wang},
  \bibinfo{person}{Michael Stonebraker}, \bibinfo{person}{Ahmed~K. Elmagarmid},
  \bibinfo{person}{Ihab~F. Ilyas}, \bibinfo{person}{Samuel Madden},
  \bibinfo{person}{Mourad Ouzzani}, {and} \bibinfo{person}{Nan Tang}.}
  \bibinfo{year}{2017}\natexlab{}.
\newblock \showarticletitle{The Data Civilizer System}. In
  \bibinfo{booktitle}{\emph{{CIDR}}}.
\newblock


\bibitem[\protect\citeauthoryear{Diaconu, Freedman, Ismert, Larson, Mittal,
  Stonecipher, Verma, and Zwilling}{Diaconu et~al\mbox{.}}{2013}]%
        {Hekaton13}
\bibfield{author}{\bibinfo{person}{Cristian Diaconu}, \bibinfo{person}{Craig
  Freedman}, \bibinfo{person}{Erik Ismert}, \bibinfo{person}{Per{-}{\AA}ke
  Larson}, \bibinfo{person}{Pravin Mittal}, \bibinfo{person}{Ryan Stonecipher},
  \bibinfo{person}{Nitin Verma}, {and} \bibinfo{person}{Mike Zwilling}.}
  \bibinfo{year}{2013}\natexlab{}.
\newblock \showarticletitle{{Hekaton: SQL Server's Memory-Optimized OLTP
  Engine}}. In \bibinfo{booktitle}{\emph{{SIGMOD}}}.
  \bibinfo{pages}{1243--1254}.
\newblock


\bibitem[\protect\citeauthoryear{Dong, Bai, Kim, Chen, Liu, and Li}{Dong
  et~al\mbox{.}}{2020}]%
        {DongBKCL020}
\bibfield{author}{\bibinfo{person}{Liming Dong}, \bibinfo{person}{Qiushi Bai},
  \bibinfo{person}{Taewoo Kim}, \bibinfo{person}{Taiji Chen},
  \bibinfo{person}{Weidong Liu}, {and} \bibinfo{person}{Chen Li}.}
  \bibinfo{year}{2020}\natexlab{}.
\newblock \showarticletitle{Marviq: Quality-Aware Geospatial Visualization of
  Range-Selection Queries Using Materialization}. In
  \bibinfo{booktitle}{\emph{{SIGMOD}}}. \bibinfo{pages}{67--82}.
\newblock


\bibitem[\protect\citeauthoryear{Doraiswamy and Freire}{Doraiswamy and
  Freire}{2020}]%
        {DBLP:conf/sigmod/DoraiswamyF20}
\bibfield{author}{\bibinfo{person}{Harish Doraiswamy} {and}
  \bibinfo{person}{Juliana Freire}.} \bibinfo{year}{2020}\natexlab{}.
\newblock \showarticletitle{A GPU-friendly Geometric Data Model and Algebra for
  Spatial Queries}. In \bibinfo{booktitle}{\emph{{SIGMOD}}}.
  \bibinfo{pages}{1875--1885}.
\newblock


\bibitem[\protect\citeauthoryear{Duggan, Elmore, Stonebraker, Balazinska, Howe,
  Kepner, Madden, Maier, Mattson, and Zdonik}{Duggan et~al\mbox{.}}{2015}]%
        {BigDAWG15}
\bibfield{author}{\bibinfo{person}{Jennie Duggan}, \bibinfo{person}{Aaron~J.
  Elmore}, \bibinfo{person}{Michael Stonebraker}, \bibinfo{person}{Magdalena
  Balazinska}, \bibinfo{person}{Bill Howe}, \bibinfo{person}{Jeremy Kepner},
  \bibinfo{person}{Sam Madden}, \bibinfo{person}{David Maier},
  \bibinfo{person}{Tim Mattson}, {and} \bibinfo{person}{Stanley~B. Zdonik}.}
  \bibinfo{year}{2015}\natexlab{}.
\newblock \showarticletitle{{The BigDAWG Polystore System}}.
\newblock \bibinfo{journal}{\emph{{SIGMOD} Record}} \bibinfo{volume}{44},
  \bibinfo{number}{2} (\bibinfo{year}{2015}), \bibinfo{pages}{11--16}.
\newblock


\bibitem[\protect\citeauthoryear{Eldawy and Mokbel}{Eldawy and Mokbel}{2015}]%
        {DBLP:conf/icde/EldawyM15}
\bibfield{author}{\bibinfo{person}{Ahmed Eldawy} {and}
  \bibinfo{person}{Mohamed~F. Mokbel}.} \bibinfo{year}{2015}\natexlab{}.
\newblock \showarticletitle{{SpatialHadoop: {A} MapReduce Framework for Spatial
  Data}}. In \bibinfo{booktitle}{\emph{{ICDE}}}. \bibinfo{pages}{1352--1363}.
\newblock


\bibitem[\protect\citeauthoryear{Eltabakh, Aref, Elmagarmid, and
  Ouzzani}{Eltabakh et~al\mbox{.}}{2014}]%
        {DBLP:journals/tkde/EltabakhAEO14}
\bibfield{author}{\bibinfo{person}{Mohamed~Y. Eltabakh},
  \bibinfo{person}{Walid~G. Aref}, \bibinfo{person}{Ahmed~K. Elmagarmid}, {and}
  \bibinfo{person}{Mourad Ouzzani}.} \bibinfo{year}{2014}\natexlab{}.
\newblock \showarticletitle{HandsOn {DB:} Managing Data Dependencies Involving
  Human Actions}.
\newblock \bibinfo{journal}{\emph{{IEEE} TKDE.}} \bibinfo{volume}{26},
  \bibinfo{number}{9} (\bibinfo{year}{2014}), \bibinfo{pages}{2193--2206}.
\newblock


\bibitem[\protect\citeauthoryear{Faerber, Kemper, Larson, Levandoski, Neumann,
  and Pavlo}{Faerber et~al\mbox{.}}{2017}]%
        {MMDB17}
\bibfield{author}{\bibinfo{person}{Franz Faerber}, \bibinfo{person}{Alfons
  Kemper}, \bibinfo{person}{Per{-}{\AA}ke Larson}, \bibinfo{person}{Justin~J.
  Levandoski}, \bibinfo{person}{Thomas Neumann}, {and} \bibinfo{person}{Andrew
  Pavlo}.} \bibinfo{year}{2017}\natexlab{}.
\newblock \showarticletitle{{Main Memory Database Systems}}.
\newblock \bibinfo{journal}{\emph{Foundations and Trends in Databases}}
  \bibinfo{volume}{8}, \bibinfo{number}{1-2} (\bibinfo{year}{2017}),
  \bibinfo{pages}{1--130}.
\newblock


\bibitem[\protect\citeauthoryear{Fernandez, Abedjan, Koko, Yuan, Madden, and
  Stonebraker}{Fernandez et~al\mbox{.}}{2018}]%
        {DBLP:conf/icde/FernandezAKYMS18}
\bibfield{author}{\bibinfo{person}{Raul~Castro Fernandez},
  \bibinfo{person}{Ziawasch Abedjan}, \bibinfo{person}{Famien Koko},
  \bibinfo{person}{Gina Yuan}, \bibinfo{person}{Samuel Madden}, {and}
  \bibinfo{person}{Michael Stonebraker}.} \bibinfo{year}{2018}\natexlab{}.
\newblock \showarticletitle{Aurum: {A} Data Discovery System}. In
  \bibinfo{booktitle}{\emph{{ICDE}}}. \bibinfo{pages}{1001--1012}.
\newblock


\bibitem[\protect\citeauthoryear{Frackowiak, Ailamaki, and Kherif}{Frackowiak
  et~al\mbox{.}}{2016}]%
        {3Dbrain}
\bibfield{author}{\bibinfo{person}{Richard Frackowiak},
  \bibinfo{person}{Anastasia Ailamaki}, {and} \bibinfo{person}{Ferath Kherif}.}
  \bibinfo{year}{2016}\natexlab{}.
\newblock \showarticletitle{Federating and Integrating What We Know About the
  Brain at All Scales: Computer Science Meets the Clinical Neurosciences}.
\newblock  (\bibinfo{year}{2016}), \bibinfo{pages}{157--170}.
\newblock


\bibitem[\protect\citeauthoryear{Geisberger, Sanders, Schultes, and
  Delling}{Geisberger et~al\mbox{.}}{2008}]%
        {CH-reference}
\bibfield{author}{\bibinfo{person}{Robert Geisberger}, \bibinfo{person}{Peter
  Sanders}, \bibinfo{person}{Dominik Schultes}, {and} \bibinfo{person}{Daniel
  Delling}.} \bibinfo{year}{2008}\natexlab{}.
\newblock \showarticletitle{Contraction Hierarchies: Faster and Simpler
  Hierarchical Routing in Road Networks}. In \bibinfo{booktitle}{\emph{WEA}}.
\newblock


\bibitem[\protect\citeauthoryear{Guo, Feng, Cong, and Bao}{Guo
  et~al\mbox{.}}{2018}]%
        {GuoFCB18}
\bibfield{author}{\bibinfo{person}{Tao Guo}, \bibinfo{person}{Kaiyu Feng},
  \bibinfo{person}{Gao Cong}, {and} \bibinfo{person}{Zhifeng Bao}.}
  \bibinfo{year}{2018}\natexlab{}.
\newblock \showarticletitle{Efficient Selection of Geospatial Data on Maps for
  Interactive and Visualized Exploration}. In
  \bibinfo{booktitle}{\emph{{SIGMOD}}}. \bibinfo{pages}{567--582}.
\newblock


\bibitem[\protect\citeauthoryear{Hassan, Kuznetsova, Jeong, Aref, and
  Sadoghi}{Hassan et~al\mbox{.}}{2018a}]%
        {GRFusion-EDBT}
\bibfield{author}{\bibinfo{person}{Mohamed~S. Hassan}, \bibinfo{person}{Tatiana
  Kuznetsova}, \bibinfo{person}{Hyun~Chai Jeong}, \bibinfo{person}{Walid~G.
  Aref}, {and} \bibinfo{person}{Mohammad Sadoghi}.}
  \bibinfo{year}{2018}\natexlab{a}.
\newblock \showarticletitle{Extending In-Memory Relational Database Engines
  with Native Graph Support}. In \bibinfo{booktitle}{\emph{{EDBT}}}.
  \bibinfo{pages}{25--36}.
\newblock


\bibitem[\protect\citeauthoryear{Hassan, Kuznetsova, Jeong, Aref, and
  Sadoghi}{Hassan et~al\mbox{.}}{2018b}]%
        {GRFusion-SIGMOD-demo}
\bibfield{author}{\bibinfo{person}{Mohamed~S. Hassan}, \bibinfo{person}{Tatiana
  Kuznetsova}, \bibinfo{person}{Hyun~Chai Jeong}, \bibinfo{person}{Walid~G.
  Aref}, {and} \bibinfo{person}{Mohammad Sadoghi}.}
  \bibinfo{year}{2018}\natexlab{b}.
\newblock \showarticletitle{GRFusion: Graphs as First-Class Citizens in
  Main-Memory Relational Database Systems}. In
  \bibinfo{booktitle}{\emph{{SIGMOD}}}. \bibinfo{pages}{1789--1792}.
\newblock


\bibitem[\protect\citeauthoryear{Hong, Jung, and Shim}{Hong
  et~al\mbox{.}}{2022}]%
        {9792226}
\bibfield{author}{\bibinfo{person}{Daeyoung Hong}, \bibinfo{person}{Woohwan
  Jung}, {and} \bibinfo{person}{Kyuseok Shim}.}
  \bibinfo{year}{2022}\natexlab{}.
\newblock \showarticletitle{Collecting Geospatial Data Under Local Differential
  Privacy With Improving Frequency Estimation}.
\newblock \bibinfo{journal}{\emph{{IEEE} TKDE}} (\bibinfo{year}{2022}),
  \bibinfo{pages}{1--12}.
\newblock


\bibitem[\protect\citeauthoryear{Hosseini, Yin, Zhang, Zhou, and
  Sadiq}{Hosseini et~al\mbox{.}}{2016}]%
        {Meihui2016}
\bibfield{author}{\bibinfo{person}{Saeid Hosseini}, \bibinfo{person}{Hongzhi
  Yin}, \bibinfo{person}{Meihui Zhang}, \bibinfo{person}{Xiaofang Zhou}, {and}
  \bibinfo{person}{Shazia Sadiq}.} \bibinfo{year}{2016}\natexlab{}.
\newblock \showarticletitle{Jointly Modeling Heterogeneous Temporal Properties
  in Location Recommendation}. In \bibinfo{booktitle}{\emph{DASFAA}}.
  \bibinfo{pages}{490--506}.
\newblock


\bibitem[\protect\citeauthoryear{Huang, Shekhar, and Xiong}{Huang
  et~al\mbox{.}}{2004}]%
        {DBLP:journals/tkde/HuangSX04}
\bibfield{author}{\bibinfo{person}{Yan Huang}, \bibinfo{person}{Shashi
  Shekhar}, {and} \bibinfo{person}{Hui Xiong}.}
  \bibinfo{year}{2004}\natexlab{}.
\newblock \showarticletitle{Discovering Colocation Patterns from Spatial Data
  Sets: {A} General Approach}.
\newblock \bibinfo{journal}{\emph{{IEEE} TKDE}} \bibinfo{volume}{16},
  \bibinfo{number}{12} (\bibinfo{year}{2004}), \bibinfo{pages}{1472--1485}.
\newblock


\bibitem[\protect\citeauthoryear{Inoue, Moriyama, Komatsu, and Nakatani}{Inoue
  et~al\mbox{.}}{2007}]%
        {DBLP:conf/IEEEpact/InoueMKN07}
\bibfield{author}{\bibinfo{person}{Hiroshi Inoue}, \bibinfo{person}{Takao
  Moriyama}, \bibinfo{person}{Hideaki Komatsu}, {and} \bibinfo{person}{Toshio
  Nakatani}.} \bibinfo{year}{2007}\natexlab{}.
\newblock \showarticletitle{AA-Sort: {A} New Parallel Sorting Algorithm for
  Multi-Core {SIMD} Processors}. In \bibinfo{booktitle}{\emph{{PACT}}}.
  \bibinfo{pages}{189--198}.
\newblock


\bibitem[\protect\citeauthoryear{Inoue and Taura}{Inoue and Taura}{2015}]%
        {DBLP:journals/pvldb/InoueT15}
\bibfield{author}{\bibinfo{person}{Hiroshi Inoue} {and}
  \bibinfo{person}{Kenjiro Taura}.} \bibinfo{year}{2015}\natexlab{}.
\newblock \showarticletitle{{SIMD-} and Cache-Friendly Algorithm for Sorting an
  Array of Structures}.
\newblock \bibinfo{journal}{\emph{{PVLDB}}} \bibinfo{volume}{8},
  \bibinfo{number}{11} (\bibinfo{year}{2015}), \bibinfo{pages}{1274--1285}.
\newblock


\bibitem[\protect\citeauthoryear{Kalamatianos, Fakas, and
  Mamoulis}{Kalamatianos et~al\mbox{.}}{2021}]%
        {KalamatianosFM21}
\bibfield{author}{\bibinfo{person}{Georgios Kalamatianos},
  \bibinfo{person}{Georgios~John Fakas}, {and} \bibinfo{person}{Nikos
  Mamoulis}.} \bibinfo{year}{2021}\natexlab{}.
\newblock \showarticletitle{Proportionality in Spatial Keyword Search}. In
  \bibinfo{booktitle}{\emph{{SIGMOD}}}. \bibinfo{pages}{885--897}.
\newblock


\bibitem[\protect\citeauthoryear{Kim, Sedlar, Chhugani, Kaldewey, Nguyen, Blas,
  Lee, Satish, and Dubey}{Kim et~al\mbox{.}}{2009}]%
        {DBLP:journals/pvldb/KimSCKNBLSD09}
\bibfield{author}{\bibinfo{person}{Changkyu Kim}, \bibinfo{person}{Eric
  Sedlar}, \bibinfo{person}{Jatin Chhugani}, \bibinfo{person}{Tim Kaldewey},
  \bibinfo{person}{Anthony~D. Nguyen}, \bibinfo{person}{Andrea~Di Blas},
  \bibinfo{person}{Victor~W. Lee}, \bibinfo{person}{Nadathur Satish}, {and}
  \bibinfo{person}{Pradeep Dubey}.} \bibinfo{year}{2009}\natexlab{}.
\newblock \showarticletitle{Sort vs. Hash Revisited: Fast Join Implementation
  on Modern Multi-Core CPUs}.
\newblock \bibinfo{journal}{\emph{{PVLDB}}} \bibinfo{volume}{2},
  \bibinfo{number}{2} (\bibinfo{year}{2009}), \bibinfo{pages}{1378--1389}.
\newblock


\bibitem[\protect\citeauthoryear{Klonatos, Koch, Rompf, and Chafi}{Klonatos
  et~al\mbox{.}}{2014}]%
        {KlonatosKRC14}
\bibfield{author}{\bibinfo{person}{Yannis Klonatos}, \bibinfo{person}{Christoph
  Koch}, \bibinfo{person}{Tiark Rompf}, {and} \bibinfo{person}{Hassan Chafi}.}
  \bibinfo{year}{2014}\natexlab{}.
\newblock \showarticletitle{Building Efficient Query Engines in a High-Level
  Language}.
\newblock \bibinfo{journal}{\emph{{PVLDB}}} \bibinfo{volume}{7},
  \bibinfo{number}{10} (\bibinfo{year}{2014}), \bibinfo{pages}{853--864}.
\newblock


\bibitem[\protect\citeauthoryear{Kraska, Beutel, Chi, Dean, and
  Polyzotis}{Kraska et~al\mbox{.}}{2018}]%
        {kraska2018case}
\bibfield{author}{\bibinfo{person}{Tim Kraska}, \bibinfo{person}{Alex Beutel},
  \bibinfo{person}{Ed~H Chi}, \bibinfo{person}{Jeffrey Dean}, {and}
  \bibinfo{person}{Neoklis Polyzotis}.} \bibinfo{year}{2018}\natexlab{}.
\newblock \showarticletitle{The case for learned index structures}. In
  \bibinfo{booktitle}{\emph{{SIGMOD}}}. \bibinfo{pages}{489--504}.
\newblock


\bibitem[\protect\citeauthoryear{Lang, Neumann, Kemper, and Boncz}{Lang
  et~al\mbox{.}}{2019}]%
        {DBLP:journals/pvldb/LangNKB19}
\bibfield{author}{\bibinfo{person}{Harald Lang}, \bibinfo{person}{Thomas
  Neumann}, \bibinfo{person}{Alfons Kemper}, {and} \bibinfo{person}{Peter~A.
  Boncz}.} \bibinfo{year}{2019}\natexlab{}.
\newblock \showarticletitle{Performance-Optimal Filtering: Bloom overtakes
  Cuckoo at High-Throughput}.
\newblock \bibinfo{journal}{\emph{{PVLDB}}} \bibinfo{volume}{12},
  \bibinfo{number}{5} (\bibinfo{year}{2019}), \bibinfo{pages}{502--515}.
\newblock


\bibitem[\protect\citeauthoryear{Larson, Clinciu, Hanson, Oks, Price,
  Rangarajan, Surna, and Zhou}{Larson et~al\mbox{.}}{2011}]%
        {DBLP:conf/sigmod/LarsonCHOPRSZ11}
\bibfield{author}{\bibinfo{person}{Per{-}{\AA}ke Larson},
  \bibinfo{person}{Cipri Clinciu}, \bibinfo{person}{Eric~N. Hanson},
  \bibinfo{person}{Artem Oks}, \bibinfo{person}{Susan~L. Price},
  \bibinfo{person}{Srikumar Rangarajan}, \bibinfo{person}{Aleksandras Surna},
  {and} \bibinfo{person}{Qingqing Zhou}.} \bibinfo{year}{2011}\natexlab{}.
\newblock \showarticletitle{{SQL} server column store indexes}. In
  \bibinfo{booktitle}{\emph{{SIGMOD}}}. \bibinfo{pages}{1177--1184}.
\newblock


\bibitem[\protect\citeauthoryear{Lee, Kwon, F{\"{a}}rber, Muehle, Lee,
  Bensberg, Lee, Lee, and Lehner}{Lee et~al\mbox{.}}{2013}]%
        {HANA13}
\bibfield{author}{\bibinfo{person}{Juchang Lee}, \bibinfo{person}{Yong~Sik
  Kwon}, \bibinfo{person}{Franz F{\"{a}}rber}, \bibinfo{person}{Michael
  Muehle}, \bibinfo{person}{Chulwon Lee}, \bibinfo{person}{Christian Bensberg},
  \bibinfo{person}{Joo{-}Yeon Lee}, \bibinfo{person}{Arthur~H. Lee}, {and}
  \bibinfo{person}{Wolfgang Lehner}.} \bibinfo{year}{2013}\natexlab{}.
\newblock \showarticletitle{{SAP {HANA} Distributed In-memory Database System:
  Transaction, Session, and Metadata Management}}. In
  \bibinfo{booktitle}{\emph{{ICDE}}}. \bibinfo{pages}{1165--1173}.
\newblock


\bibitem[\protect\citeauthoryear{Leis, Boncz, Kemper, and Neumann}{Leis
  et~al\mbox{.}}{2014}]%
        {DBLP:conf/sigmod/LeisBK014}
\bibfield{author}{\bibinfo{person}{Viktor Leis}, \bibinfo{person}{Peter~A.
  Boncz}, \bibinfo{person}{Alfons Kemper}, {and} \bibinfo{person}{Thomas
  Neumann}.} \bibinfo{year}{2014}\natexlab{}.
\newblock \showarticletitle{Morsel-driven parallelism: a NUMA-aware query
  evaluation framework for the many-core age}. In
  \bibinfo{booktitle}{\emph{{SIGMOD}}}. \bibinfo{pages}{743--754}.
\newblock


\bibitem[\protect\citeauthoryear{Levandoski, Lomet, and Sengupta}{Levandoski
  et~al\mbox{.}}{2013}]%
        {LevandoskiLS13a}
\bibfield{author}{\bibinfo{person}{Justin~J. Levandoski},
  \bibinfo{person}{David~B. Lomet}, {and} \bibinfo{person}{Sudipta Sengupta}.}
  \bibinfo{year}{2013}\natexlab{}.
\newblock \showarticletitle{The Bw-Tree: {A} B-tree for new hardware
  platforms}. In \bibinfo{booktitle}{\emph{{ICDE}}}. \bibinfo{pages}{302--313}.
\newblock


\bibitem[\protect\citeauthoryear{Li, Lu, Zheng, Yang, and Pan}{Li
  et~al\mbox{.}}{2020}]%
        {Li0ZY020}
\bibfield{author}{\bibinfo{person}{Pengfei Li}, \bibinfo{person}{Hua Lu},
  \bibinfo{person}{Qian Zheng}, \bibinfo{person}{Long Yang}, {and}
  \bibinfo{person}{Gang Pan}.} \bibinfo{year}{2020}\natexlab{}.
\newblock \showarticletitle{{LISA:} {A} Learned Index Structure for Spatial
  Data}. In \bibinfo{booktitle}{\emph{{SIGMOD}}}. \bibinfo{pages}{2119--2133}.
\newblock


\bibitem[\protect\citeauthoryear{Li, Pandis, M{\"{u}}ller, Raman, and
  Lohman}{Li et~al\mbox{.}}{2013}]%
        {DBLP:conf/cidr/LiPMRL13}
\bibfield{author}{\bibinfo{person}{Yinan Li}, \bibinfo{person}{Ippokratis
  Pandis}, \bibinfo{person}{Ren{\'{e}} M{\"{u}}ller},
  \bibinfo{person}{Vijayshankar Raman}, {and} \bibinfo{person}{Guy~M. Lohman}.}
  \bibinfo{year}{2013}\natexlab{}.
\newblock \showarticletitle{NUMA-aware algorithms: the case of data shuffling}.
  In \bibinfo{booktitle}{\emph{{CIDR}}}.
\newblock


\bibitem[\protect\citeauthoryear{Liakos, Papakonstantinopoulou, and
  Kotidis}{Liakos et~al\mbox{.}}{2022}]%
        {Chimp22}
\bibfield{author}{\bibinfo{person}{Panagiotis Liakos}, \bibinfo{person}{Katia
  Papakonstantinopoulou}, {and} \bibinfo{person}{Yannis Kotidis}.}
  \bibinfo{year}{2022}\natexlab{}.
\newblock \showarticletitle{Chimp: Efficient Lossless Floating Point
  Compression for Time Series Databases}.
\newblock \bibinfo{journal}{\emph{{PVLDB}}}  \bibinfo{volume}{15}
  (\bibinfo{year}{2022}).
\newblock


\bibitem[\protect\citeauthoryear{Lin and Chen}{Lin and Chen}{2008}]%
        {Lin4568473}
\bibfield{author}{\bibinfo{person}{Hung-Yi Lin} {and}
  \bibinfo{person}{Shih-Ying Chen}.} \bibinfo{year}{2008}\natexlab{}.
\newblock \showarticletitle{High Indexing Compression for Spatial Databases}.
  In \bibinfo{booktitle}{\emph{{CIT} Workshops}}. \bibinfo{pages}{20--25}.
\newblock


\bibitem[\protect\citeauthoryear{Lomet}{Lomet}{2001}]%
        {Lomet01}
\bibfield{author}{\bibinfo{person}{David~B. Lomet}.}
  \bibinfo{year}{2001}\natexlab{}.
\newblock \showarticletitle{The Evolution of Effective B-tree: Page
  Organization and Techniques: {A} Personal Account}.
\newblock \bibinfo{journal}{\emph{{SIGMOD} Rec.}} \bibinfo{volume}{30},
  \bibinfo{number}{3} (\bibinfo{year}{2001}), \bibinfo{pages}{64--69}.
\newblock


\bibitem[\protect\citeauthoryear{Lou, Zhang, Zheng, Xie, Wang, and Huang}{Lou
  et~al\mbox{.}}{2009}]%
        {LouZZXWH09}
\bibfield{author}{\bibinfo{person}{Yin Lou}, \bibinfo{person}{Chengyang Zhang},
  \bibinfo{person}{Yu Zheng}, \bibinfo{person}{Xing Xie}, \bibinfo{person}{Wei
  Wang}, {and} \bibinfo{person}{Yan Huang}.} \bibinfo{year}{2009}\natexlab{}.
\newblock \showarticletitle{Map-matching for low-sampling-rate {GPS}
  trajectories}. In \bibinfo{booktitle}{\emph{{SIGSPATIAL}}}.
  \bibinfo{pages}{352--361}.
\newblock


\bibitem[\protect\citeauthoryear{Lu, Ding, Lo, Minhas, and Wang}{Lu
  et~al\mbox{.}}{2021}]%
        {LuDLMW21}
\bibfield{author}{\bibinfo{person}{Baotong Lu}, \bibinfo{person}{Jialin Ding},
  \bibinfo{person}{Eric Lo}, \bibinfo{person}{Umar~Farooq Minhas}, {and}
  \bibinfo{person}{Tianzheng Wang}.} \bibinfo{year}{2021}\natexlab{}.
\newblock \showarticletitle{{APEX:} {A} High-Performance Learned Index on
  Persistent Memory}.
\newblock \bibinfo{journal}{\emph{{PVLDB}}} \bibinfo{volume}{15},
  \bibinfo{number}{3} (\bibinfo{year}{2021}), \bibinfo{pages}{597--610}.
\newblock


\bibitem[\protect\citeauthoryear{Lu, Hao, Wang, and Lo}{Lu
  et~al\mbox{.}}{2020}]%
        {LuHWL20}
\bibfield{author}{\bibinfo{person}{Baotong Lu}, \bibinfo{person}{Xiangpeng
  Hao}, \bibinfo{person}{Tianzheng Wang}, {and} \bibinfo{person}{Eric Lo}.}
  \bibinfo{year}{2020}\natexlab{}.
\newblock \showarticletitle{Dash: Scalable Hashing on Persistent Memory}.
\newblock \bibinfo{journal}{\emph{{PVLDB}}} \bibinfo{volume}{13},
  \bibinfo{number}{8} (\bibinfo{year}{2020}), \bibinfo{pages}{1147--1161}.
\newblock


\bibitem[\protect\citeauthoryear{Lu and Holubov{\'{a}}}{Lu and
  Holubov{\'{a}}}{2019}]%
        {MMDBSurvey19}
\bibfield{author}{\bibinfo{person}{Jiaheng Lu} {and} \bibinfo{person}{Irena
  Holubov{\'{a}}}.} \bibinfo{year}{2019}\natexlab{}.
\newblock \showarticletitle{{Multi-model Databases: {A} New Journey to Handle
  the Variety of Data}}.
\newblock \bibinfo{journal}{\emph{ACM Computing Surveys (CSUR)}}
  \bibinfo{volume}{52}, \bibinfo{number}{3} (\bibinfo{year}{2019}),
  \bibinfo{pages}{55:1--55:38}.
\newblock


\bibitem[\protect\citeauthoryear{Mahmood, Aly, Qadah, Rezig, Daghistani,
  Madkour, Abdelhamid, Hassan, Aref, and Basalamah}{Mahmood
  et~al\mbox{.}}{2015}]%
        {Tornado-reference}
\bibfield{author}{\bibinfo{person}{Ahmed~R. Mahmood}, \bibinfo{person}{Ahmed~M.
  Aly}, \bibinfo{person}{Thamir Qadah}, \bibinfo{person}{El~Kindi Rezig},
  \bibinfo{person}{Anas Daghistani}, \bibinfo{person}{Amgad Madkour},
  \bibinfo{person}{Ahmed~S. Abdelhamid}, \bibinfo{person}{Mohamed~S. Hassan},
  \bibinfo{person}{Walid~G. Aref}, {and} \bibinfo{person}{Saleh~M. Basalamah}.}
  \bibinfo{year}{2015}\natexlab{}.
\newblock \showarticletitle{Tornado: {A} Distributed Spatio-Textual Stream
  Processing System}.
\newblock \bibinfo{journal}{\emph{{PVLDB}}} \bibinfo{volume}{8},
  \bibinfo{number}{12} (\bibinfo{year}{2015}), \bibinfo{pages}{2020--2023}.
\newblock


\bibitem[\protect\citeauthoryear{Mokbel, Alarabi, Bao, Eldawy, Magdy, Sarwat,
  Waytas, and Yackel}{Mokbel et~al\mbox{.}}{2013}]%
        {10.1007/978-3-642-40235-7_3}
\bibfield{author}{\bibinfo{person}{Mohamed~F. Mokbel}, \bibinfo{person}{Louai
  Alarabi}, \bibinfo{person}{Jie Bao}, \bibinfo{person}{Ahmed Eldawy},
  \bibinfo{person}{Amr Magdy}, \bibinfo{person}{Mohamed Sarwat},
  \bibinfo{person}{Ethan Waytas}, {and} \bibinfo{person}{Steven Yackel}.}
  \bibinfo{year}{2013}\natexlab{}.
\newblock \showarticletitle{MNTG: An Extensible Web-Based Traffic Generator}.
  In \bibinfo{booktitle}{\emph{SSTD}}. \bibinfo{pages}{38--55}.
\newblock


\bibitem[\protect\citeauthoryear{Neumann}{Neumann}{2011}]%
        {Neumann2011}
\bibfield{author}{\bibinfo{person}{T. Neumann}.}
  \bibinfo{year}{2011}\natexlab{}.
\newblock \showarticletitle{Efficiently Compiling Efficient Query Plans for
  Modern Hardware}.
\newblock \bibinfo{journal}{\emph{{PVLDB}}} \bibinfo{volume}{4},
  \bibinfo{number}{9} (\bibinfo{year}{2011}), \bibinfo{pages}{539–550}.
\newblock


\bibitem[\protect\citeauthoryear{Newson and Krumm}{Newson and Krumm}{2009}]%
        {NewsonK09}
\bibfield{author}{\bibinfo{person}{Paul Newson} {and} \bibinfo{person}{John
  Krumm}.} \bibinfo{year}{2009}\natexlab{}.
\newblock \showarticletitle{Hidden Markov map matching through noise and
  sparseness}. In \bibinfo{booktitle}{\emph{{SIGSPATIAL}}}.
  \bibinfo{pages}{336--343}.
\newblock


\bibitem[\protect\citeauthoryear{O'Neil, Cheng, Gawlick, and O'Neil}{O'Neil
  et~al\mbox{.}}{1996}]%
        {DBLP:journals/acta/ONeilCGO96}
\bibfield{author}{\bibinfo{person}{Patrick~E. O'Neil}, \bibinfo{person}{Edward
  Cheng}, \bibinfo{person}{Dieter Gawlick}, {and} \bibinfo{person}{Elizabeth~J.
  O'Neil}.} \bibinfo{year}{1996}\natexlab{}.
\newblock \showarticletitle{The Log-Structured Merge-Tree (LSM-Tree)}.
\newblock \bibinfo{journal}{\emph{Acta Informatica}} \bibinfo{volume}{33},
  \bibinfo{number}{4} (\bibinfo{year}{1996}), \bibinfo{pages}{351--385}.
\newblock


\bibitem[\protect\citeauthoryear{Osogami and Raymond}{Osogami and
  Raymond}{2013}]%
        {OsogamiR13}
\bibfield{author}{\bibinfo{person}{Takayuki Osogami} {and}
  \bibinfo{person}{Rudy Raymond}.} \bibinfo{year}{2013}\natexlab{}.
\newblock \showarticletitle{Map Matching with Inverse Reinforcement Learning}.
  In \bibinfo{booktitle}{\emph{{IJCAI}}}. \bibinfo{pages}{2547--2553}.
\newblock


\bibitem[\protect\citeauthoryear{Papadakis, Mandilaras, Mamoulis, and
  Koubarakis}{Papadakis et~al\mbox{.}}{2022}]%
        {10.1145/3510025}
\bibfield{author}{\bibinfo{person}{George Papadakis}, \bibinfo{person}{George
  Mandilaras}, \bibinfo{person}{Nikos Mamoulis}, {and} \bibinfo{person}{Manolis
  Koubarakis}.} \bibinfo{year}{2022}\natexlab{}.
\newblock \showarticletitle{Static and Dynamic Progressive Geospatial
  Interlinking}.
\newblock \bibinfo{journal}{\emph{{ACM} TSAS}} \bibinfo{volume}{8},
  \bibinfo{number}{2} (\bibinfo{year}{2022}).
\newblock


\bibitem[\protect\citeauthoryear{Patel, Deshmukh, Zhu, Potti, Zhang, Spehlmann,
  Memisoglu, and Saurabh}{Patel et~al\mbox{.}}{2018a}]%
        {DBLP:journals/pvldb/PatelDZPZSMS18}
\bibfield{author}{\bibinfo{person}{Jignesh~M. Patel}, \bibinfo{person}{Harshad
  Deshmukh}, \bibinfo{person}{Jianqiao Zhu}, \bibinfo{person}{Navneet Potti},
  \bibinfo{person}{Zuyu Zhang}, \bibinfo{person}{Marc Spehlmann},
  \bibinfo{person}{Hakan Memisoglu}, {and} \bibinfo{person}{Saket Saurabh}.}
  \bibinfo{year}{2018}\natexlab{a}.
\newblock \showarticletitle{Quickstep: {A} Data Platform Based on the
  Scaling-Up Approach}.
\newblock \bibinfo{journal}{\emph{{PVLDB}}} \bibinfo{volume}{11},
  \bibinfo{number}{6} (\bibinfo{year}{2018}), \bibinfo{pages}{663--676}.
\newblock


\bibitem[\protect\citeauthoryear{Patel, Deshmukh, Zhu, Potti, Zhang, Spehlmann,
  Memisoglu, and Saurabh}{Patel et~al\mbox{.}}{2018b}]%
        {PatelDZPZSMS18}
\bibfield{author}{\bibinfo{person}{Jignesh~M. Patel}, \bibinfo{person}{Harshad
  Deshmukh}, \bibinfo{person}{Jianqiao Zhu}, \bibinfo{person}{Navneet Potti},
  \bibinfo{person}{Zuyu Zhang}, \bibinfo{person}{Marc Spehlmann},
  \bibinfo{person}{Hakan Memisoglu}, {and} \bibinfo{person}{Saket Saurabh}.}
  \bibinfo{year}{2018}\natexlab{b}.
\newblock \showarticletitle{Quickstep: {A} Data Platform Based on the
  Scaling-Up Approach}.
\newblock \bibinfo{journal}{\emph{{PVLDB}}} \bibinfo{volume}{11},
  \bibinfo{number}{6} (\bibinfo{year}{2018}), \bibinfo{pages}{663--676}.
\newblock


\bibitem[\protect\citeauthoryear{Pelkonen, Franklin, Cavallaro, Huang, Meza,
  Teller, and Veeraraghavan}{Pelkonen et~al\mbox{.}}{2015}]%
        {PelkonenFCHMTV15}
\bibfield{author}{\bibinfo{person}{Tuomas Pelkonen}, \bibinfo{person}{Scott
  Franklin}, \bibinfo{person}{Paul Cavallaro}, \bibinfo{person}{Qi Huang},
  \bibinfo{person}{Justin Meza}, \bibinfo{person}{Justin Teller}, {and}
  \bibinfo{person}{Kaushik Veeraraghavan}.} \bibinfo{year}{2015}\natexlab{}.
\newblock \showarticletitle{Gorilla: {A} Fast, Scalable, In-Memory Time Series
  Database}.
\newblock \bibinfo{journal}{\emph{{PVLDB}}} \bibinfo{volume}{8},
  \bibinfo{number}{12} (\bibinfo{year}{2015}), \bibinfo{pages}{1816--1827}.
\newblock


\bibitem[\protect\citeauthoryear{Peng and Samet}{Peng and Samet}{2015}]%
        {do-analytical-reference}
\bibfield{author}{\bibinfo{person}{Shangfu Peng} {and} \bibinfo{person}{Hanan
  Samet}.} \bibinfo{year}{2015}\natexlab{}.
\newblock \showarticletitle{Analytical Queries on Road Networks: An
  Experimental Evaluation of Two System Architectures}. In
  \bibinfo{booktitle}{\emph{SIGSPATIAL}}.
\newblock


\bibitem[\protect\citeauthoryear{Polychroniou and Ross}{Polychroniou and
  Ross}{2013}]%
        {DBLP:conf/damon/PolychroniouR13}
\bibfield{author}{\bibinfo{person}{Orestis Polychroniou} {and}
  \bibinfo{person}{Kenneth~A. Ross}.} \bibinfo{year}{2013}\natexlab{}.
\newblock \showarticletitle{High throughput heavy hitter aggregation for modern
  {SIMD} processors}. In \bibinfo{booktitle}{\emph{{DaMoN}}}.
  \bibinfo{pages}{6}.
\newblock


\bibitem[\protect\citeauthoryear{Polychroniou and Ross}{Polychroniou and
  Ross}{2014}]%
        {DBLP:conf/sigmod/PolychroniouR14}
\bibfield{author}{\bibinfo{person}{Orestis Polychroniou} {and}
  \bibinfo{person}{Kenneth~A. Ross}.} \bibinfo{year}{2014}\natexlab{}.
\newblock \showarticletitle{A comprehensive study of main-memory partitioning
  and its application to large-scale comparison- and radix-sort}. In
  \bibinfo{booktitle}{\emph{{SIGMOD}}}. \bibinfo{pages}{755--766}.
\newblock


\bibitem[\protect\citeauthoryear{Polychroniou and Ross}{Polychroniou and
  Ross}{2015}]%
        {DBLP:conf/damon/PolychroniouR15}
\bibfield{author}{\bibinfo{person}{Orestis Polychroniou} {and}
  \bibinfo{person}{Kenneth~A. Ross}.} \bibinfo{year}{2015}\natexlab{}.
\newblock \showarticletitle{Efficient Lightweight Compression Alongside Fast
  Scans}. In \bibinfo{booktitle}{\emph{{DaMoN}}}. \bibinfo{pages}{9:1--9:6}.
\newblock


\bibitem[\protect\citeauthoryear{Polychroniou and Ross}{Polychroniou and
  Ross}{2019}]%
        {DBLP:conf/damon/PolychroniouR19}
\bibfield{author}{\bibinfo{person}{Orestis Polychroniou} {and}
  \bibinfo{person}{Kenneth~A. Ross}.} \bibinfo{year}{2019}\natexlab{}.
\newblock \showarticletitle{Towards Practical Vectorized Analytical Query
  Engines}. In \bibinfo{booktitle}{\emph{{DaMoN}}}.
  \bibinfo{pages}{10:1--10:7}.
\newblock


\bibitem[\protect\citeauthoryear{Psaroudakis, Scheuer, May, Sellami, and
  Ailamaki}{Psaroudakis et~al\mbox{.}}{2016}]%
        {DBLP:journals/pvldb/PsaroudakisSMSA16}
\bibfield{author}{\bibinfo{person}{Iraklis Psaroudakis},
  \bibinfo{person}{Tobias Scheuer}, \bibinfo{person}{Norman May},
  \bibinfo{person}{Abdelkader Sellami}, {and} \bibinfo{person}{Anastasia
  Ailamaki}.} \bibinfo{year}{2016}\natexlab{}.
\newblock \showarticletitle{Adaptive NUMA-aware data placement and task
  scheduling for analytical workloads in main-memory column-stores}.
\newblock \bibinfo{journal}{\emph{{PVLDB}}} \bibinfo{volume}{10},
  \bibinfo{number}{2} (\bibinfo{year}{2016}), \bibinfo{pages}{37--48}.
\newblock


\bibitem[\protect\citeauthoryear{Qi, Liu, Jensen, and Kulik}{Qi
  et~al\mbox{.}}{2020}]%
        {qi13effectively}
\bibfield{author}{\bibinfo{person}{Jianzhong Qi}, \bibinfo{person}{Guanli Liu},
  \bibinfo{person}{Christian~S Jensen}, {and} \bibinfo{person}{Lars Kulik}.}
  \bibinfo{year}{2020}\natexlab{}.
\newblock \showarticletitle{Effectively learning spatial indices}.
\newblock \bibinfo{journal}{\emph{{PVLDB}}} \bibinfo{volume}{13},
  \bibinfo{number}{12} (\bibinfo{year}{2020}), \bibinfo{pages}{2341--2354}.
\newblock


\bibitem[\protect\citeauthoryear{Raman, Attaluri, Barber, Chainani, Kalmuk,
  KulandaiSamy, Leenstra, Lightstone, Liu, Lohman, Malkemus, M{\"{u}}ller,
  Pandis, Schiefer, Sharpe, Sidle, Storm, and Zhang}{Raman
  et~al\mbox{.}}{2013}]%
        {DBLP:journals/pvldb/RamanABCKKLLLLMMPSSSSZ13}
\bibfield{author}{\bibinfo{person}{Vijayshankar Raman},
  \bibinfo{person}{Gopi~K. Attaluri}, \bibinfo{person}{Ronald Barber},
  \bibinfo{person}{Naresh Chainani}, \bibinfo{person}{David Kalmuk},
  \bibinfo{person}{Vincent KulandaiSamy}, \bibinfo{person}{Jens Leenstra},
  \bibinfo{person}{Sam Lightstone}, \bibinfo{person}{Shaorong Liu},
  \bibinfo{person}{Guy~M. Lohman}, \bibinfo{person}{Tim Malkemus},
  \bibinfo{person}{Ren{\'{e}} M{\"{u}}ller}, \bibinfo{person}{Ippokratis
  Pandis}, \bibinfo{person}{Berni Schiefer}, \bibinfo{person}{David Sharpe},
  \bibinfo{person}{Richard Sidle}, \bibinfo{person}{Adam~J. Storm}, {and}
  \bibinfo{person}{Liping Zhang}.} \bibinfo{year}{2013}\natexlab{}.
\newblock \showarticletitle{{DB2} with {BLU} Acceleration: So Much More than
  Just a Column Store}.
\newblock \bibinfo{journal}{\emph{{PVLDB}}} \bibinfo{volume}{6},
  \bibinfo{number}{11} (\bibinfo{year}{2013}), \bibinfo{pages}{1080--1091}.
\newblock


\bibitem[\protect\citeauthoryear{Sanders and Schultes}{Sanders and
  Schultes}{2005}]%
        {HH05-reference}
\bibfield{author}{\bibinfo{person}{Peter Sanders} {and}
  \bibinfo{person}{Dominik Schultes}.} \bibinfo{year}{2005}\natexlab{}.
\newblock \showarticletitle{Highway Hierarchies Hasten Exact Shortest Path
  Queries}. In \bibinfo{booktitle}{\emph{{ESA}}}. \bibinfo{pages}{568--579}.
\newblock


\bibitem[\protect\citeauthoryear{Sankaranarayanan and Samet}{Sankaranarayanan
  and Samet}{2009}]%
        {do2009-reference}
\bibfield{author}{\bibinfo{person}{Jagan Sankaranarayanan} {and}
  \bibinfo{person}{Hanan Samet}.} \bibinfo{year}{2009}\natexlab{}.
\newblock \showarticletitle{Distance Oracles for Spatial Networks}. In
  \bibinfo{booktitle}{\emph{{ICDE}}}. \bibinfo{pages}{652--663}.
\newblock


\bibitem[\protect\citeauthoryear{Satish, Kim, Chhugani, Nguyen, Lee, Kim, and
  Dubey}{Satish et~al\mbox{.}}{2010}]%
        {DBLP:conf/sigmod/SatishKCNLKD10}
\bibfield{author}{\bibinfo{person}{Nadathur Satish}, \bibinfo{person}{Changkyu
  Kim}, \bibinfo{person}{Jatin Chhugani}, \bibinfo{person}{Anthony~D. Nguyen},
  \bibinfo{person}{Victor~W. Lee}, \bibinfo{person}{Daehyun Kim}, {and}
  \bibinfo{person}{Pradeep Dubey}.} \bibinfo{year}{2010}\natexlab{}.
\newblock \showarticletitle{Fast sort on CPUs and GPUs: a case for bandwidth
  oblivious {SIMD} sort}. In \bibinfo{booktitle}{\emph{{SIGMOD}}}.
  \bibinfo{pages}{351--362}.
\newblock


\bibitem[\protect\citeauthoryear{Sethi, Traverso, Sundstrom, Phillips, Xie,
  Sun, Yegitbasi, Jin, Hwang, Shingte, and Berner}{Sethi et~al\mbox{.}}{2019}]%
        {DBLP:conf/icde/SethiTSPXSYJHSB19}
\bibfield{author}{\bibinfo{person}{Raghav Sethi}, \bibinfo{person}{Martin
  Traverso}, \bibinfo{person}{Dain Sundstrom}, \bibinfo{person}{David
  Phillips}, \bibinfo{person}{Wenlei Xie}, \bibinfo{person}{Yutian Sun},
  \bibinfo{person}{Nezih Yegitbasi}, \bibinfo{person}{Haozhun Jin},
  \bibinfo{person}{Eric Hwang}, \bibinfo{person}{Nileema Shingte}, {and}
  \bibinfo{person}{Christopher Berner}.} \bibinfo{year}{2019}\natexlab{}.
\newblock \showarticletitle{Presto: {SQL} on Everything}. In
  \bibinfo{booktitle}{\emph{{ICDE}}}. \bibinfo{pages}{1802--1813}.
\newblock


\bibitem[\protect\citeauthoryear{Shaham, Ghinita, Ahuja, Krumm, and
  Shahabi}{Shaham et~al\mbox{.}}{2021}]%
        {10.1145/3474717.3483943}
\bibfield{author}{\bibinfo{person}{Sina Shaham}, \bibinfo{person}{Gabriel
  Ghinita}, \bibinfo{person}{Ritesh Ahuja}, \bibinfo{person}{John Krumm}, {and}
  \bibinfo{person}{Cyrus Shahabi}.} \bibinfo{year}{2021}\natexlab{}.
\newblock \showarticletitle{HTF: Homogeneous Tree Framework for
  Differentially-Private Release of Location Data}. In
  \bibinfo{booktitle}{\emph{{SIGSPATIAL}}}. \bibinfo{pages}{184–194}.
\newblock


\bibitem[\protect\citeauthoryear{Shekhar, Lu, and Zhang}{Shekhar
  et~al\mbox{.}}{2003}]%
        {DBLP:journals/geoinformatica/ShekharLZ03}
\bibfield{author}{\bibinfo{person}{Shashi Shekhar},
  \bibinfo{person}{Chang{-}Tien Lu}, {and} \bibinfo{person}{Pusheng Zhang}.}
  \bibinfo{year}{2003}\natexlab{}.
\newblock \showarticletitle{A Unified Approach to Detecting Spatial Outliers}.
\newblock \bibinfo{journal}{\emph{GeoInformatica}} \bibinfo{volume}{7},
  \bibinfo{number}{2} (\bibinfo{year}{2003}), \bibinfo{pages}{139--166}.
\newblock


\bibitem[\protect\citeauthoryear{Shin, Wang, and Aref}{Shin
  et~al\mbox{.}}{2021}]%
        {Shin0A21}
\bibfield{author}{\bibinfo{person}{Jaewoo Shin}, \bibinfo{person}{Jianguo
  Wang}, {and} \bibinfo{person}{Walid~G. Aref}.}
  \bibinfo{year}{2021}\natexlab{}.
\newblock \showarticletitle{The {LSM} RUM-Tree: {A} Log Structured Merge R-Tree
  for Update-intensive Spatial Workloads}. In \bibinfo{booktitle}{\emph{ICDE}}.
  \bibinfo{pages}{2285--2290}.
\newblock


\bibitem[\protect\citeauthoryear{Shute, Vingralek, Samwel, Handy, Whipkey,
  Rollins, Oancea, Littlefield, Menestrina, Ellner, Cieslewicz, Rae, Stancescu,
  and Apte}{Shute et~al\mbox{.}}{2013}]%
        {DBLP:journals/pvldb/ShuteVSHWROLMECRSA13}
\bibfield{author}{\bibinfo{person}{Jeff Shute}, \bibinfo{person}{Radek
  Vingralek}, \bibinfo{person}{Bart Samwel}, \bibinfo{person}{Ben Handy},
  \bibinfo{person}{Chad Whipkey}, \bibinfo{person}{Eric Rollins},
  \bibinfo{person}{Mircea Oancea}, \bibinfo{person}{Kyle Littlefield},
  \bibinfo{person}{David Menestrina}, \bibinfo{person}{Stephan Ellner},
  \bibinfo{person}{John Cieslewicz}, \bibinfo{person}{Ian Rae},
  \bibinfo{person}{Traian Stancescu}, {and} \bibinfo{person}{Himani Apte}.}
  \bibinfo{year}{2013}\natexlab{}.
\newblock \showarticletitle{{F1:} {A} Distributed {SQL} Database That Scales}.
\newblock \bibinfo{journal}{\emph{{PVLDB}}} \bibinfo{volume}{6},
  \bibinfo{number}{11} (\bibinfo{year}{2013}), \bibinfo{pages}{1068--1079}.
\newblock


\bibitem[\protect\citeauthoryear{Shvachko, Kuang, Radia, and Chansler}{Shvachko
  et~al\mbox{.}}{2010}]%
        {hadoop-reference}
\bibfield{author}{\bibinfo{person}{Konstantin Shvachko},
  \bibinfo{person}{Hairong Kuang}, \bibinfo{person}{Sanjay Radia}, {and}
  \bibinfo{person}{Robert Chansler}.} \bibinfo{year}{2010}\natexlab{}.
\newblock \showarticletitle{The Hadoop Distributed File System}. In
  \bibinfo{booktitle}{\emph{MSST}}. \bibinfo{pages}{1--10}.
\newblock


\bibitem[\protect\citeauthoryear{Silva, Aref, Larson, Pearson, and Ali}{Silva
  et~al\mbox{.}}{2013}]%
        {DBLP:journals/vldb/SilvaALPA13}
\bibfield{author}{\bibinfo{person}{Yasin~N. Silva}, \bibinfo{person}{Walid~G.
  Aref}, \bibinfo{person}{Per{-}{\AA}ke Larson}, \bibinfo{person}{Spencer
  Pearson}, {and} \bibinfo{person}{Mohamed~H. Ali}.}
  \bibinfo{year}{2013}\natexlab{}.
\newblock \showarticletitle{Similarity queries: their conceptual evaluation,
  transformations, and processing}.
\newblock \bibinfo{journal}{\emph{{VLDB} J.}} \bibinfo{volume}{22},
  \bibinfo{number}{3} (\bibinfo{year}{2013}), \bibinfo{pages}{395--420}.
\newblock


\bibitem[\protect\citeauthoryear{Silva, Xiong, and Aref}{Silva
  et~al\mbox{.}}{2009}]%
        {DBLP:journals/vldb/SilvaXA09}
\bibfield{author}{\bibinfo{person}{Yasin~N. Silva}, \bibinfo{person}{Xiaopeng
  Xiong}, {and} \bibinfo{person}{Walid~G. Aref}.}
  \bibinfo{year}{2009}\natexlab{}.
\newblock \showarticletitle{The RUM-tree: supporting frequent updates in
  R-trees using memos}.
\newblock \bibinfo{journal}{\emph{{VLDB} J.}} \bibinfo{volume}{18},
  \bibinfo{number}{3} (\bibinfo{year}{2009}), \bibinfo{pages}{719--738}.
\newblock


\bibitem[\protect\citeauthoryear{Song, Kim, and Yoo}{Song
  et~al\mbox{.}}{2004}]%
        {SongKY04}
\bibfield{author}{\bibinfo{person}{Seok~Il Song}, \bibinfo{person}{Young~Ho
  Kim}, {and} \bibinfo{person}{Jae~Soo Yoo}.} \bibinfo{year}{2004}\natexlab{}.
\newblock \showarticletitle{An Enhanced Concurrency Control Scheme for
  Multidimensional Index Structures}.
\newblock \bibinfo{journal}{\emph{{IEEE} TKDE}} \bibinfo{volume}{16},
  \bibinfo{number}{1} (\bibinfo{year}{2004}), \bibinfo{pages}{97--111}.
\newblock


\bibitem[\protect\citeauthoryear{Stonebraker and Rowe}{Stonebraker and
  Rowe}{1986}]%
        {postgres-reference}
\bibfield{author}{\bibinfo{person}{Michael Stonebraker} {and}
  \bibinfo{person}{Lawrence~A. Rowe}.} \bibinfo{year}{1986}\natexlab{}.
\newblock \showarticletitle{The Design of POSTGRES}. In
  \bibinfo{booktitle}{\emph{{SIGMOD}}}. \bibinfo{pages}{340–355}.
\newblock


\bibitem[\protect\citeauthoryear{Stonebraker and Weisberg}{Stonebraker and
  Weisberg}{2013}]%
        {voltdb-reference}
\bibfield{author}{\bibinfo{person}{Michael Stonebraker} {and}
  \bibinfo{person}{Ariel Weisberg}.} \bibinfo{year}{2013}\natexlab{}.
\newblock \showarticletitle{The VoltDB Main Memory DBMS}.
\newblock \bibinfo{journal}{\emph{IEEE Data Eng. Bull.}}  \bibinfo{volume}{36}
  (\bibinfo{year}{2013}), \bibinfo{pages}{21--27}.
\newblock


\bibitem[\protect\citeauthoryear{Tahboub and Rompf}{Tahboub and Rompf}{2016}]%
        {TahboubR16}
\bibfield{author}{\bibinfo{person}{Ruby~Y. Tahboub} {and}
  \bibinfo{person}{Tiark Rompf}.} \bibinfo{year}{2016}\natexlab{}.
\newblock \showarticletitle{On supporting compilation in spatial query engines:
  (vision paper)}. In \bibinfo{booktitle}{\emph{{SIGSPATIAL}}}.
  \bibinfo{pages}{9:1--9:4}.
\newblock


\bibitem[\protect\citeauthoryear{Tang, Yu, Malluhi, Ouzzani, and Aref}{Tang
  et~al\mbox{.}}{2016}]%
        {locationSparkReference}
\bibfield{author}{\bibinfo{person}{MingJie Tang}, \bibinfo{person}{Yongyang
  Yu}, \bibinfo{person}{Qutaibah~M. Malluhi}, \bibinfo{person}{Mourad Ouzzani},
  {and} \bibinfo{person}{Walid~G. Aref}.} \bibinfo{year}{2016}\natexlab{}.
\newblock \showarticletitle{LocationSpark: {A} Distributed In-Memory Data
  Management System for Big Spatial Data}.
\newblock \bibinfo{journal}{\emph{{PVLDB}}} \bibinfo{volume}{9},
  \bibinfo{number}{13} (\bibinfo{year}{2016}), \bibinfo{pages}{1565--1568}.
\newblock


\bibitem[\protect\citeauthoryear{Tao, Liu, Wang, Battle, Demiralp, Chang, and
  Stonebraker}{Tao et~al\mbox{.}}{2019}]%
        {Kyrix2019}
\bibfield{author}{\bibinfo{person}{Wenbo Tao}, \bibinfo{person}{Xiaoyu Liu},
  \bibinfo{person}{Yedi Wang}, \bibinfo{person}{Leilani Battle},
  \bibinfo{person}{{\c{C}}agatay Demiralp}, \bibinfo{person}{Remco Chang},
  {and} \bibinfo{person}{Michael Stonebraker}.}
  \bibinfo{year}{2019}\natexlab{}.
\newblock \showarticletitle{Kyrix: Interactive Pan/Zoom Visualizations at
  Scale}.
\newblock \bibinfo{journal}{\emph{Comput. Graph. Forum}} \bibinfo{volume}{38},
  \bibinfo{number}{3} (\bibinfo{year}{2019}), \bibinfo{pages}{529--540}.
\newblock


\bibitem[\protect\citeauthoryear{Terrizzano, Schwarz, Roth, and
  Colino}{Terrizzano et~al\mbox{.}}{2015}]%
        {DBLP:conf/cidr/TerrizzanoSRC15}
\bibfield{author}{\bibinfo{person}{Ignacio~G. Terrizzano},
  \bibinfo{person}{Peter~M. Schwarz}, \bibinfo{person}{Mary Roth}, {and}
  \bibinfo{person}{John~E. Colino}.} \bibinfo{year}{2015}\natexlab{}.
\newblock \showarticletitle{Data Wrangling: The Challenging Yourney from the
  Wild to the Lake}. In \bibinfo{booktitle}{\emph{{CIDR}}}.
\newblock


\bibitem[\protect\citeauthoryear{To, Shahabi, and Xiong}{To
  et~al\mbox{.}}{2018}]%
        {8509301}
\bibfield{author}{\bibinfo{person}{Hien To}, \bibinfo{person}{Cyrus Shahabi},
  {and} \bibinfo{person}{Li Xiong}.} \bibinfo{year}{2018}\natexlab{}.
\newblock \showarticletitle{Privacy-Preserving Online Task Assignment in
  Spatial Crowdsourcing with Untrusted Server}. In
  \bibinfo{booktitle}{\emph{{ICDE}}}. \bibinfo{pages}{833--844}.
\newblock


\bibitem[\protect\citeauthoryear{Tsitsigkos, Bouros, Mamoulis, and
  Terrovitis}{Tsitsigkos et~al\mbox{.}}{2019}]%
        {10.1145/3347146.3359343}
\bibfield{author}{\bibinfo{person}{Dimitrios Tsitsigkos},
  \bibinfo{person}{Panagiotis Bouros}, \bibinfo{person}{Nikos Mamoulis}, {and}
  \bibinfo{person}{Manolis Terrovitis}.} \bibinfo{year}{2019}\natexlab{}.
\newblock \showarticletitle{Parallel In-Memory Evaluation of Spatial Joins}. In
  \bibinfo{booktitle}{\emph{SIGSPATIAL}}. \bibinfo{pages}{516–519}.
\newblock


\bibitem[\protect\citeauthoryear{Vu, Belussi, Migliorini, and Eldawy}{Vu
  et~al\mbox{.}}{2021}]%
        {10.1145/3474717.3484217}
\bibfield{author}{\bibinfo{person}{Tin Vu}, \bibinfo{person}{Alberto Belussi},
  \bibinfo{person}{Sara Migliorini}, {and} \bibinfo{person}{Ahmed Eldawy}.}
  \bibinfo{year}{2021}\natexlab{}.
\newblock \showarticletitle{A Learned Query Optimizer for Spatial Join}. In
  \bibinfo{booktitle}{\emph{{SIGSPATIAL}}}. \bibinfo{pages}{458–467}.
\newblock


\bibitem[\protect\citeauthoryear{Wang, Wang, Idreos, {\"{O}}zsu, and Aref}{Wang
  et~al\mbox{.}}{2022}]%
        {DSM22}
\bibfield{author}{\bibinfo{person}{Ruihong Wang}, \bibinfo{person}{Jianguo
  Wang}, \bibinfo{person}{Stratos Idreos}, \bibinfo{person}{M.~Tamer
  {\"{O}}zsu}, {and} \bibinfo{person}{Walid~G. Aref}.}
  \bibinfo{year}{2022}\natexlab{}.
\newblock \showarticletitle{The Case for Distributed Shared-Memory Databases
  with RDMA-Enabled Memory Disaggregation}.
\newblock \bibinfo{journal}{\emph{CoRR}}  \bibinfo{volume}{abs/2207.03027}
  (\bibinfo{year}{2022}).
\newblock


\bibitem[\protect\citeauthoryear{Willhalm, Popovici, Boshmaf, Plattner, Zeier,
  and Schaffner}{Willhalm et~al\mbox{.}}{2009}]%
        {DBLP:journals/pvldb/WillhalmPBPZS09}
\bibfield{author}{\bibinfo{person}{Thomas Willhalm}, \bibinfo{person}{Nicolae
  Popovici}, \bibinfo{person}{Yazan Boshmaf}, \bibinfo{person}{Hasso Plattner},
  \bibinfo{person}{Alexander Zeier}, {and} \bibinfo{person}{Jan Schaffner}.}
  \bibinfo{year}{2009}\natexlab{}.
\newblock \showarticletitle{SIMD-Scan: Ultra Fast in-Memory Table Scan using
  on-Chip Vector Processing Units}.
\newblock \bibinfo{journal}{\emph{{PVLDB}}} \bibinfo{volume}{2},
  \bibinfo{number}{1} (\bibinfo{year}{2009}), \bibinfo{pages}{385--394}.
\newblock


\bibitem[\protect\citeauthoryear{Xie, Li, Yao, Li, Zhou, and Guo}{Xie
  et~al\mbox{.}}{2016}]%
        {XieL0LZG16}
\bibfield{author}{\bibinfo{person}{Dong Xie}, \bibinfo{person}{Feifei Li},
  \bibinfo{person}{Bin Yao}, \bibinfo{person}{Gefei Li}, \bibinfo{person}{Liang
  Zhou}, {and} \bibinfo{person}{Minyi Guo}.} \bibinfo{year}{2016}\natexlab{}.
\newblock \showarticletitle{Simba: Efficient In-Memory Spatial Analytics}. In
  \bibinfo{booktitle}{\emph{{SIGMOD}}}. \bibinfo{pages}{1071--1085}.
\newblock


\bibitem[\protect\citeauthoryear{Ye, Ross, and Vesdapunt}{Ye
  et~al\mbox{.}}{2011}]%
        {DBLP:conf/damon/YeRV11}
\bibfield{author}{\bibinfo{person}{Yang Ye}, \bibinfo{person}{Kenneth~A. Ross},
  {and} \bibinfo{person}{Norases Vesdapunt}.} \bibinfo{year}{2011}\natexlab{}.
\newblock \showarticletitle{Scalable aggregation on multicore processors}. In
  \bibinfo{booktitle}{\emph{{DaMoN}}}. \bibinfo{pages}{1--9}.
\newblock


\bibitem[\protect\citeauthoryear{Yoo, Shekhar, and Celik}{Yoo
  et~al\mbox{.}}{2005}]%
        {DBLP:conf/icdm/YooSC05}
\bibfield{author}{\bibinfo{person}{Jin~Soung Yoo}, \bibinfo{person}{Shashi
  Shekhar}, {and} \bibinfo{person}{Mete Celik}.}
  \bibinfo{year}{2005}\natexlab{}.
\newblock \showarticletitle{A Join-Less Approach for Co-Location Pattern
  Mining: {A} Summary of Results}. In \bibinfo{booktitle}{\emph{{ICDM}}}.
  \bibinfo{pages}{813--816}.
\newblock


\bibitem[\protect\citeauthoryear{Yu, Wu, and Sarwat}{Yu et~al\mbox{.}}{2015}]%
        {GeoSpark}
\bibfield{author}{\bibinfo{person}{Jia Yu}, \bibinfo{person}{Jinxuan Wu}, {and}
  \bibinfo{person}{Mohamed Sarwat}.} \bibinfo{year}{2015}\natexlab{}.
\newblock \showarticletitle{GeoSpark: a cluster computing framework for
  processing large-scale spatial data}. In
  \bibinfo{booktitle}{\emph{{SIGSPATIAL}}}. \bibinfo{pages}{70:1--70:4}.
\newblock


\bibitem[\protect\citeauthoryear{Yu, Bezerra, Pavlo, Devadas, and
  Stonebraker}{Yu et~al\mbox{.}}{2014}]%
        {YuBPDS14}
\bibfield{author}{\bibinfo{person}{Xiangyao Yu}, \bibinfo{person}{George
  Bezerra}, \bibinfo{person}{Andrew Pavlo}, \bibinfo{person}{Srinivas Devadas},
  {and} \bibinfo{person}{Michael Stonebraker}.}
  \bibinfo{year}{2014}\natexlab{}.
\newblock \showarticletitle{Staring into the Abyss: An Evaluation of
  Concurrency Control with One Thousand Cores}.
\newblock \bibinfo{journal}{\emph{{PVLDB}}} \bibinfo{volume}{8},
  \bibinfo{number}{3} (\bibinfo{year}{2014}), \bibinfo{pages}{209--220}.
\newblock


\bibitem[\protect\citeauthoryear{Yuan, Zheng, Zhang, Xie, and Sun}{Yuan
  et~al\mbox{.}}{2010}]%
        {YuanZZXS10}
\bibfield{author}{\bibinfo{person}{Jing Yuan}, \bibinfo{person}{Yu Zheng},
  \bibinfo{person}{Chengyang Zhang}, \bibinfo{person}{Xing Xie}, {and}
  \bibinfo{person}{Guangzhong Sun}.} \bibinfo{year}{2010}\natexlab{}.
\newblock \showarticletitle{An Interactive-Voting Based Map Matching
  Algorithm}. In \bibinfo{booktitle}{\emph{{MDM}}}. \bibinfo{pages}{43--52}.
\newblock


\bibitem[\protect\citeauthoryear{Zaharia, Chowdhury, Franklin, Shenker, and
  Stoica}{Zaharia et~al\mbox{.}}{2010}]%
        {spark-reference}
\bibfield{author}{\bibinfo{person}{Matei Zaharia}, \bibinfo{person}{Mosharaf
  Chowdhury}, \bibinfo{person}{Michael~J. Franklin}, \bibinfo{person}{Scott
  Shenker}, {and} \bibinfo{person}{Ion Stoica}.}
  \bibinfo{year}{2010}\natexlab{}.
\newblock \showarticletitle{Spark: Cluster Computing with Working Sets}. In
  \bibinfo{booktitle}{\emph{{HotCloud}}}.
\newblock


\bibitem[\protect\citeauthoryear{Zeighami, Ahuja, Ghinita, and
  Shahabi}{Zeighami et~al\mbox{.}}{2022}]%
        {10.14778/3510397.3510404}
\bibfield{author}{\bibinfo{person}{Sepanta Zeighami}, \bibinfo{person}{Ritesh
  Ahuja}, \bibinfo{person}{Gabriel Ghinita}, {and} \bibinfo{person}{Cyrus
  Shahabi}.} \bibinfo{year}{2022}\natexlab{}.
\newblock \showarticletitle{A Neural Database for Differentially Private
  Spatial Range Queries}.
\newblock \bibinfo{journal}{\emph{{PVLDB}}} \bibinfo{volume}{15},
  \bibinfo{number}{5} (\bibinfo{year}{2022}), \bibinfo{pages}{1066–1078}.
\newblock


\bibitem[\protect\citeauthoryear{Zhou and Ross}{Zhou and Ross}{2002}]%
        {DBLP:conf/sigmod/ZhouR02}
\bibfield{author}{\bibinfo{person}{Jingren Zhou} {and}
  \bibinfo{person}{Kenneth~A. Ross}.} \bibinfo{year}{2002}\natexlab{}.
\newblock \showarticletitle{Implementing database operations using {SIMD}
  instructions}. In \bibinfo{booktitle}{\emph{{SIGMOD}}}.
  \bibinfo{pages}{145--156}.
\newblock


\end{thebibliography}
\end{sloppypar}

\end{document}